\documentclass[11pt]{JHEP3}
\usepackage{epsfig}
\usepackage{amsmath,axodraw,graphics}
\def\be{\begin{equation}}
\def\ee{\end{equation}}
\def\bea{\begin{array}}
\def\eea{\end{array}}
\def\beqa{\begin{eqnarray}}
\def\eeqa{\end{eqnarray}}
\def\beqas{\begin{eqnarray*}}
\def\eeqas{\end{eqnarray*}}

\def\bp{\begin{picture}}
\def\ep{\end{picture}}
\def\bc{\begin{center}}
\def\ec{\end{center}}
\def\bfig{\begin{figure}}
\def\efig{\end{figure}}

\def\bit{\begin{itemize}}
\def\eit{\end{itemize}}
\def\nn{\nonumber}
\def\f{\frac}

\def\[{\left[}
\def\]{\right]}
\def\({\left(}
\def\){\right)}

\def\..{\left.}
\def\.{\right.}
\def\tl{\tilde}
\def\ra{\rightarrow}
\def\la{\leftarrow}

\def\dif{\partial}
\def\tm{\times}

\def\ks{k\hspace{-0.2cm}\slash}

\def\lks{k\hspace{-0.13cm}\slash}
\def\lps{p\hspace{-0.10cm}\slash}

\def\da{\dagger}
\def\LL{{\cal L}}
\def\DD{{\cal D}}

\def\la{\lambda}

\def\al{\alpha}
\def\bt{\beta}
\def\ka{\kappa}
\def\ze{\zeta}
\def\si{\sigma}
\def\ep{\epsilon}
\def\rh{\rho}
\def\et{\eta}

\def\ga{\gamma}
\def\pa{\partial}
\def\pr{\prime}

\def\sz{\small}
\def\nz{\normalsize}
\title{Bounds on Higgs And Top Quark Masses From The Other Degenerate Vacua Near The Planck Scale With Gravitational Contributions}
\author{ Fei Wang$^{1,2}$, Guo-Li Liu$^{1,2}$, Kun Wu$^1$\\

$^1$ Department of Physics and Engineering, Zhengzhou University, Henan 450001, P. R. China\\
$^2$ State Key Laboratory of Theoretical Physics, Institute of Theoretical Physics China, Chinese Academy of Sciences, Beijing 100190, P. R. China
}

\abstract{ Based on the weak coupling expansion of gravity, we calculate the gravitational contributions to yukawa coupling, scalar quartic coupling as well as gauge couplings with general Landau-DeWitt gauge-fixing choice and a gauge preserving (of SM gauge group) cut off regularization scheme. We find that the results depend on the Landau-DeWitt gauge-fixing parameter. Based on the two loop RGE of SM couplings with one loop full gravitational contributions in harmonic gauge, we study the constraints on the higgs and top quark mass from the requirement of existing the other degenerate vacua at the Planck-dominated region. Our numerical calculations show that nature will not develop the other degenerate vacua at the Planck-dominated region with current higgs and top quark masses. On the other hand, requiring the existence of the other degenerate vacua at the Planck-dominated region will constrain the higgs and top mass to lie at approximately 130 and 174 GeV, respectively.
}

\begin{document}
\maketitle \indent
\newpage
\section{Introduction}
   The discovery of a 125 GeV Standard Model-like higgs boson by both the ATLAS and CMS collaboration \cite{atlas,cms} of the Large Hadron Collider (LHC) fills the last missing piece of the Standard Model (SM) of particle physics. Experimental data on such scalar particle agree quite well with the SM predictions and no signs of new physics beyond the SM have been observed so far.

   Although the SM seems very successful in describing the real world, there are still many theoretical and aesthetical problems in SM, such as the dark matter puzzle and the hierarchy problem that related to the existence of a fundamental higgs scalar. It also seems problematic to extent the validity range of SM to Planck scale because the renormalization group equation (RGE) running of quartic coupling $\la$ with the current higgs and top quark mass will drive $\la$ negative at large field value which could lead to another local minimum at large field value.  If such new minimum lies below our electroweak(EW) vacuum, quantum tunneling effects from the (false) EW vacuum to the (true) deeper vacuum could result in the false vacuum decay and thus the EW vacuum instability. In fact, absolute stability of the higgs potential is excluded at 98\% C.L. for $M_h < 126 {\rm GeV}$\cite{higgsbound,higgsbound2,lindner}. On the other hand, typical calculations on the tunneling rate with the central value of higgs mass indicates that the false EW vacuum is a metastable vacuum with a lifetime longer the age of the universe\cite{metastablebound}.

  A mysterious in the extrapolation of $\la$ is its slow running at high energy which is due to a combination of two factors: the decreasing behavior of all SM couplings at high energy and the nearly vanishing of $\beta_\la$ at a scale of about $10^{17}$ to $10^{18}$ GeV. Previous SM based calculations indicate that the quartic coupling $\la$ and its beta function $\beta_\la$ nearly vanish at the Planck scale. This may indicate the "multiple point criticality principle" (MPCP)\cite{mpcp} which assumes the other degenerate vacua at the Planck scale and was used to predict the top mass being $173\pm 5$ GeV and a Higgs mass $135\pm9$ GeV  in 1990s. However, the inputs of \cite{mpcp} are out of date and the calculations neglect the possible gravitational contributions which could be important near the Planck scale. On the other hand, asymptotic safety of gravity\cite{Shaposhnikov} indicate that the MPCP may arise at a typical "transition scale" $k_{tr}$ that near $M_{Pl}$. So it is meaningful to require the MPCP to be hold at some energy scale near the $M_{Pl}$ instead of exactly at $M_{Pl}$. We calculate the gravitational contributions to the beta functions of the standard model yukawa, gauge and quatic couplings to see the status of MPCP after the higgs discovery.

  The quantum effects of gravity, which is non-renormalizable by perturbative methods, can be studied in an effective theory approach and may play an important role near the Planck scale. It is interesting to note that the higgs mass had already been predicted to be 126 GeV in the fundamental theory composed of SM and the asymptotic safety of gravity before LHC discovery\cite{Shaposhnikov}. Although gravitational effects decouple in most of the discussions related to standard model, such effects can change the RGE running behavior of quartic coupling near the Planck scale. An interesting consequence of gravitational effects is the asymptotic free behavior of all gauge couplings near Planck scale when new power-law running gravitational contributions become dominant\cite{wilczek}.

  Calculation of gravitational contributions to gauge couplings by \cite{wilczek} with background field method was found by \cite{gaugedependent1,gaugedependent2} to be gauge dependent and claimed that the true contribution vanishes.
Further studies by\cite{ylwu,tom,hongjian-he,kiritsis,daum} confirms the non-zero gravitational contributions to the running of gauge couplings in \cite{wilczek}. Relevant calculations are also given with Vilkovisky¨CDeWitt effective action approach or other methods with specified gauge fixing condition\cite{scalar0,scalar1,scalar2,scalar3,artur,sunyi}. Our calculations are based on the traditional Feynman diagram methods and use a  gauge invariance preserving cut-off regularization scheme\cite{ylwu1,cut}. We should note that the weak coupling expansion of gravity used in this paper is not sufficient. Contributions from the non-renormalizable aspects of gravity could be important. The inclusion of certain higher-dimensional operators in higgs sector could possibly modify the behavior of the potential near the Planck scale \cite{scalarNR1,scalarNR2}. Typical constraints from MPCP at the Planck scale on higgs and top quark masses taking into account the gravitational contributions was given in \cite{haba}. However, the adopted gravitational contributions to yukawa couplings which is very crucial in determining the UV behavior of $\la$ do not agree in sign with our calculations. Besides, it is more preferable to require MPCP in the Planck scale dominated region instead of exactly at the Planck scale. So it is very interesting to study the status of MPCP at the Planck scale dominated region with our fully independent calculations.

    This paper is organized as follows. In sec-2, we perform the calculation of gravitational contributions to the beta function of SM yukawa, gauge and quatic coupling with the general Landau-DeWitt gauge-fixing  choice. In sec-3, we discuss the constraints from MPCP in the Planck scale dominated region based on our calculated gravitational contributions. Sec-4 contains our conclusion.

 \section{Gravitional Corrections to Yukawa Coupling}
 Quantum gravity is well known to be non-renormalizable. However, the quantum effects of gravity can be taken into account in an effective theory approach \cite{donoghue}. Physical predictions for gravitational effects are justified if we interest in physics at the energy scale $E\lesssim M_{Pl}$. Such predictions coincide with the results given by the underlying fundamental theory whatever its nature.

   The action $S$ containing gauge, scalar and fermion fields can be written as
 \begin{eqnarray}
\label{action}
 S &=&\int d^dx\sqrt{-g}\[\kappa^{-2}R
 -\f{1}{2\ze}\(\dif_\nu h^{\mu\nu}-\f{1}{2}\dif^\mu h\)^2
 + \LL_\phi +\LL_{g}+\LL_\psi-y_i\bar{\psi}_i\psi_i \phi +\cdots \],\nn
 \\
 \LL_\phi &=&
 \frac{1}{2}g^{\mu\nu}\dif_{\mu}\phi\dif_{\nu}{\phi}
 -\frac{1}{2}m^2\phi^2
 -\frac{1}{4!}{\lambda}\phi^4,\nn\\
 \LL_{\psi+g} &=& i\bar{\psi}e_a^\mu \gamma^a (D_\mu+\f{1}{8}\omega^{ab}_\mu[\gamma_a,\gamma_b]) \psi-\f{1}{4}F_{\mu\nu}^aF^{a\mu\nu},~
 \end{eqnarray}
 where $'R'$ is the Ricci scalar curvature. The vierbein ${e}_\mu^{\alpha}$ in (\ref{action}) represents
the square root of the metric ${e}_\mu^{\alpha}{e}_\nu^{\beta}\eta_{\alpha\beta}\equiv g_{\mu\nu}$ and ${e}^{\alpha\mu}\equiv {e}^{\alpha}_{\
\rho}g^{\rho\mu}$. We can rewrite $\sqrt{-g}=|{e}_\mu^{\ \alpha}|\equiv {e}$ and $\gamma^\mu={e}^{\mu}_{\
\alpha}\gamma^{\alpha}$. The spin connection can be expressed in terms of vierbein as
\beqa
\omega_\mu^{IJ}&=&\f{1}{2}e^{\nu[I}\(e^{J]}_{\nu,\mu}-e^{J]}_{\mu,\nu}+e^{J]\rho}e^{K}_\mu e_{\nu,\rho K}\),\nn\\
&=&\f{1}{2}e^{\nu I}\(\pa_\mu e_\nu^J-\pa_\nu e^J_\mu\)-\f{1}{2}e^{\nu J}\(\pa_\mu e_\nu^I-\pa_\nu e^I_\mu\)+\f{1}{2}e^{\nu I}e^{\sigma J}(\pa_\sigma e_{\nu K}-\pa_\nu e_{\sigma K}) e^K_\mu~.
\eeqa

 The reduced gravitational coupling $\,\ka =\sqrt{16\pi G_N}\approx (1.69 \times 10^{18} {\rm GeV})^{-1}$
are determined by the Planck scale given by $\,{M}_{\rm Pl}=G_N^{-\f{1}{2}}\simeq 1.220937\times 10^{19}\,$GeV.\, The weak-coupling expansion for the Einstein gravity on Minkowski metric are given by
 \beqa
 g_{\mu\nu} &=& \eta_{\mu\nu} +\ka h_{\mu\nu} \,, \nn\\
 g^{\mu\nu} &=& \eta^{\mu\nu} -\ka h^{\mu\nu}
                 +\ka^2 h^{\mu\si}h^{\nu}_{\si} +O(\ka^3) \,,\nn\\
 \sqrt{-g} &=& 1+\f{\ka}{2}h
                +\f{\ka^2}{8}(h^2-2h^{\mu\nu}h_{\mu\nu})
                +O(\ka^3),
 \eeqa
 where
 \,$\eta_{\mu\nu}=\eta^{\mu\nu}=(1,-1,-1,-1)$\, and
 \,$h=h^{\mu\nu}\eta_{\mu\nu}=h^\mu_\mu$\,.\,
The vierbein can also be expanded in term of $\ka$
\beqa
{e}_\mu^{\ \alpha}&=&\delta_\mu^{\ \alpha}+\frac{\kappa}{2}h_\mu^{\
\alpha}-\frac{\kappa^2}{8}h_{\mu\rho}h^{\rho\alpha}+\cdots\nonumber\\
{e}_{\mu \alpha}&=&{e}_\mu^{\
\beta}\eta_{\beta\alpha}=\eta_{\mu\alpha}+\frac{\kappa}{2}h_{\mu\alpha}
-\frac{\kappa^2}{8}h_{\mu\rho}h^{\rho}_{\ \alpha}+\cdots\nonumber\\
{e}^{\mu \alpha}&=&g^{\mu\nu}{e}_\nu^{\
\alpha}=\eta^{\mu\alpha}-\frac{\kappa}{2}h^{\mu\alpha}
+\frac{3\kappa^2}{8}h^{\mu\rho}h_{\rho}^{\ \alpha}+\cdots\nonumber\\
{e}^{\mu}_{\
\alpha}&=&{e}^{\mu\beta}\eta_{\beta\alpha}=\delta^{\mu}_{\
\alpha}-\frac{\kappa}{2}h^{\mu}_{\ \alpha}
+\frac{3\kappa^2}{8}h^{\mu\rho}h_{\rho\alpha}+\cdots \ .
\eeqa
From the action, we can derive the free graviton propagator,
 \beqa
 \DD_{\mu\nu,\si\rh}(p) &=& \f{i}{2(p^2+i\ep)} \[\eta_{\mu\si}\eta_{\nu\rh} +
 \eta_{\mu\rh}\eta_{\nu\si} -
 \f{2}{d-2}\et_{\mu\nu}\et_{\si\rh} \right.
 \nn\\[3mm]
 &&\hspace*{15mm} \left.
 - \f{1-\ze}{p^2}\(
 p_\mu p_\si \et_{\nu\rh} +
 p_\mu p_\rh \et_{\nu\si} +
 p_\nu p_\si \et_{\mu\rh} +
 p_\nu p_\rh \et_{\mu\si}
 \)
 \]
 \eeqa
 where $d=4$ is the space-time dimension.
 In order to calculate the gravitational contributions to the beta function of the yukawa coupling, we must calculate the fermion, scalar self energy corrections as well as the yukawa vertex corrections.

\subsection{Fermion Self Energy and The Ward-Takahashi Identity in QED}
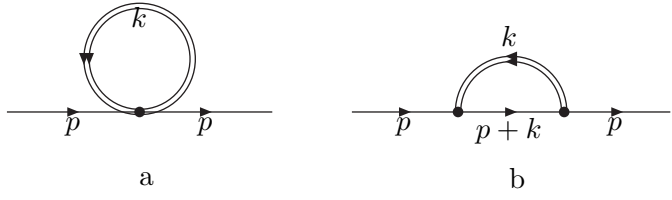
\begin{figure}[bthp]
\begin{center}
\begin{picture}(290,90)(0,0)
\Vertex(80,40){2}
\ArrowLine(30,40)(80,40)\ArrowLine(80,40)(130,40)
\ArrowArc(80,60)(19,0,360)\ArrowArc(80,60)(21,0,360)
\Text(55,38)[t]{$p$}\Text(105,38)[t]{$p$}
\Text(80,80)[t]{$k$}
\Text(80,15)[l]{a}
\Vertex(200,40){2} \Vertex(240,40){2}
\ArrowLine(160,40)(200,40)\Text(180,38)[t]{$p$}
\ArrowLine(200,40)(240,40)\Text(220,38)[t]{$p+k$}
\ArrowLine(240,40)(280,40)\Text(260,38)[t]{$p$}
\ArrowArc(220,40)(19,0,180)\Text(220,65)[b]{$k$}
\ArrowArc(220,40)(21,0,180)
\Text(220,15)[l]{b}
\end{picture}
\end{center}
\caption{The gravitational contributions to fermion self energy.}
\label{figa}
\end{figure}

Relevant diagrams that contribute to fermion self energy are given in fig.\ref{figa}.
The relevant lengthy Feynman diagrams are derived and given in the appendix.  The contribution from panel(a) is
\beqa
i\Sigma_a(\not{p})&=&\f{1}{2} \int_k D^{\mu\nu,\rho\sigma}_4[\bar{\psi}(p)\psi(p)h(k) h(-k)]\DD_{\mu\nu,\si\rh}(k)~,\nn\\
 &=&-\f{\ka^2}{16}\int_k \f{1}{k^2(k^2+i\epsilon)}\left[(3d-2)(\zeta-1) (k\cdot p)\not{k}-k^2[(3d^2-11d+7)+(4d-3)\zeta]\not{p}\right]~,\nn\\
 &=&-\f{\ka^2}{16}\int_k \f{1}{k^2+i\epsilon}\[\f{3d-2}{2}(\zeta-1)-(3d^2-11d+7)-(4d-3)\zeta\]\not{p},~\nn\\
 &=&i\ka^2\f{2+\zeta}{2}\not{p} I_2,
 \eeqa
The integral $I_2$ is defined as
\beqa
I_2=-i\int d^dk\f{1}{k^2+i\epsilon}=-\int\limits_{0}^\Lambda d^d k_E\f{1}{k_E^2}=-\f{\Lambda^2}{16\pi^2}~,
\eeqa
The contribution from panel(b) is
\beqa
i\Sigma_a(\not{p})&=& \int_k D^{\mu\nu}_3[\bar{\psi}(p)\psi(p+k)h(-k)]\f{i}{\not{p}+\not{k}-m+i\epsilon} D^{\rho\si}_3[\bar{\psi}(p+k)\psi(p)h(k)]\DD_{\mu\nu,\si\rh}(k)~,\nn\\
 &=&\f{\ka^2}{128}\int_k \f{-4\left[3(\zeta+1)k^2\not{k}+(17-3\zeta)k^2\not{p}+(18\zeta-2)(k\cdot p)\not{k}\right]}{[(p+k)^2-m^2+i\epsilon](k^2+i\epsilon)}~,\nn\\
 &=&-i\f{\ka^2}{32}(9\zeta-1+17-3\zeta-3\zeta-3)\not{p}I_2,~\nn\\
 &=&-i{\ka^2}(\f{13}{32}+\f{3}{32}\zeta)\not{p}I_2,
\eeqa

So the sum of gravitational contributions to  the self energy for fermions are given by
\beqa
i\Sigma(\not{p})=i\kappa^2(\f{19}{32}+\f{13}{32}\zeta)\not{p} I_2.
\eeqa
Then the renormalization constant $Z_\psi$ is given as
\beqa
\delta_2\equiv Z_{\psi}-1=-\ka^2(\f{19}{32}+\f{13}{32}\zeta)I_2,
\eeqa

In our renormalizaiton, we use the known consistency\cite{ylwu1,cut} cut off regularization rule
\beqa
I_{2,\mu,\nu}\equiv\int \f{d^4k}{(2\pi)^4}\f{k_\mu k_\nu}{(k^2+\Delta)^2}=\f{1}{2}\int \f{d^4k}{(2\pi)^4} \f{g_{\mu\nu}}{(k^2+\Delta)}\equiv\f{1}{2}g_{\mu\nu} I_{2,0}~.
\eeqa
 Note that the coefficient is $1/2$ for quadratic divergence while be $1/4$ for logarithm divergence. With this rule, the gauge invariance can be checked to be preserved. It is well known that gauge invariance requires that the fermion self energy counter term is equal to the fermion-antifermion-photon vertex counter term in QED. So for theoretical consistency,  it is also important to check that the previously used renormalization procedure will preserve the Ward-Takahashi identity when gravity is taken into account.

\begin{figure}[htbp]
  \begin{minipage}[t]{0.5\linewidth}
    \centering
    \includegraphics[width=3in]{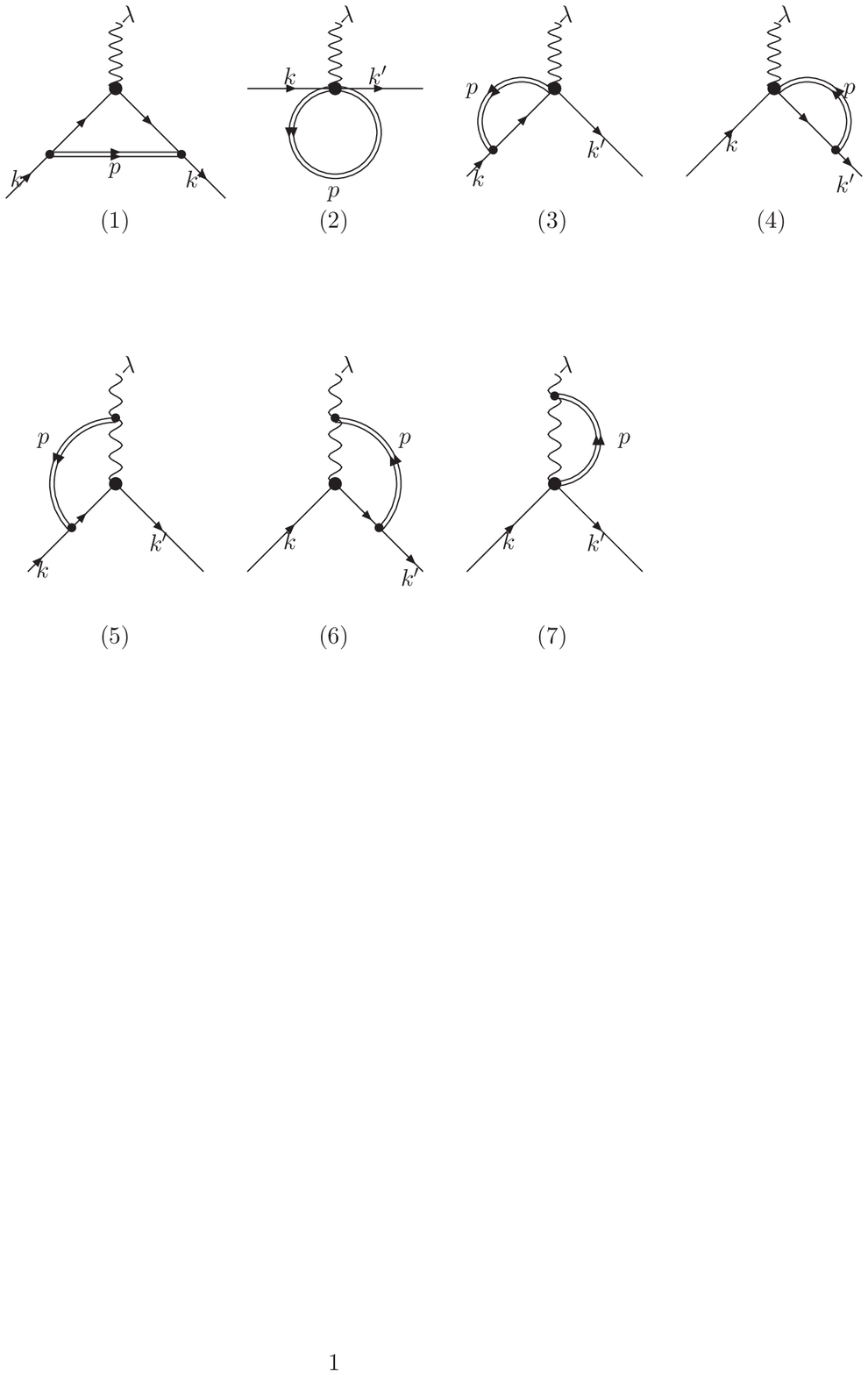}
  \end{minipage}
  \begin{minipage}[t]{0.5\linewidth}
    \centering
    \includegraphics[width=3in]{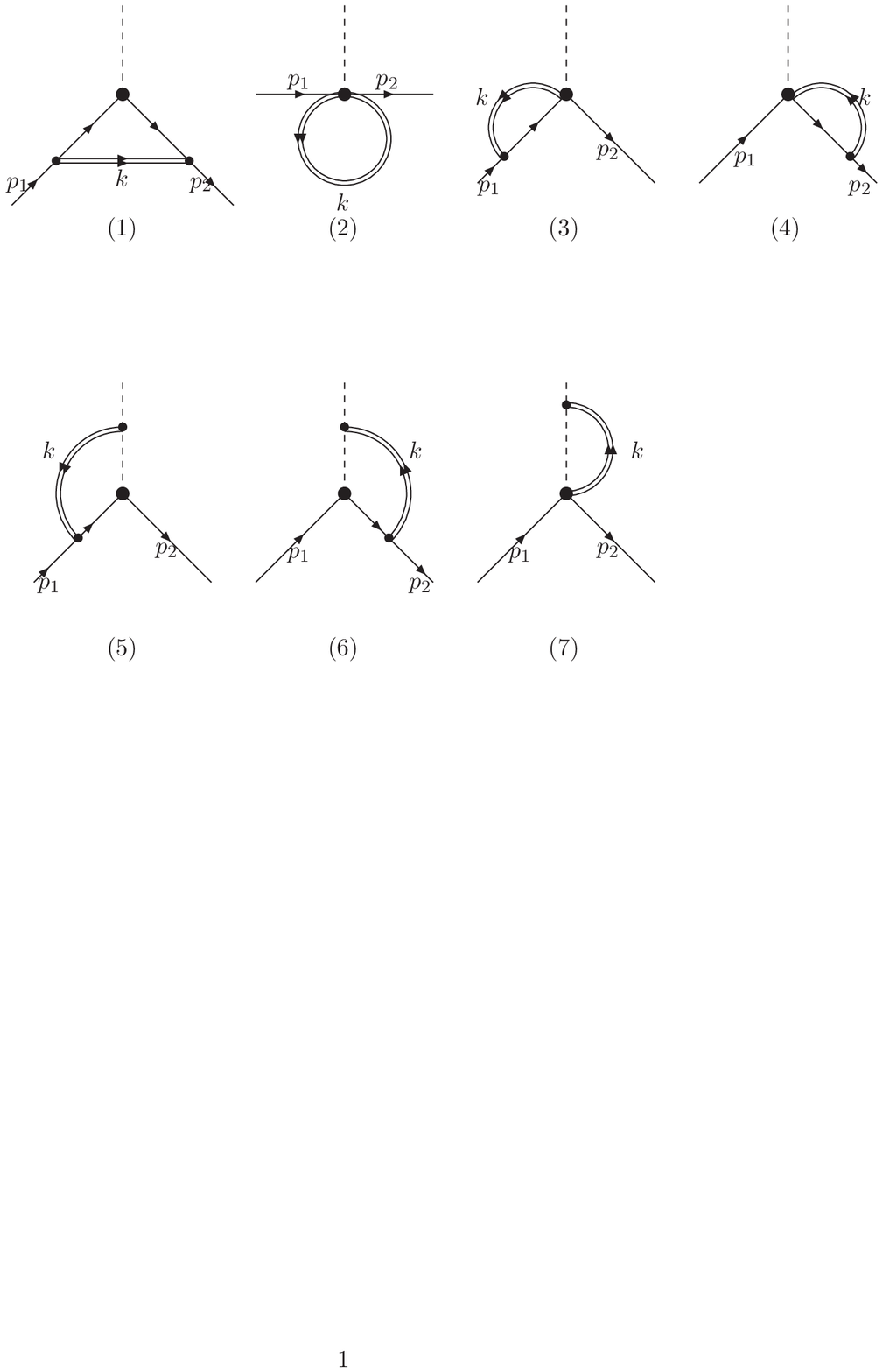}
  \end{minipage}
\caption{The gravitational contributions to the vertex corrections for gauge (left panel) and yukawa couplings (right panel).
Diagrams in the second line give no contributions because they are logarithm divergent.}
\label{fig2}
\end{figure}

Diagrams that contribute to the gauge vertex are listed in the left panel of fig.\ref{fig2} which are calculated to be

\bit
\item  Diagram (1):
\small\beqa
-i\Gamma_e^{(1)}&=&\int_p D^{\rho\sigma}_3[\bar{\psi}(k^\pr)\psi(p+k^\pr)h(p)]\f{i(\not{p}+\not{k^\pr}+m)}{(p+k^\pr)^2-m^2+i\epsilon}(-ie\gamma^\la)
\f{i(\not{p}+\not{k}+m)}{(p+k)^2-m^2+i\epsilon}\nn\\
&& ~~~~D^{\mu\nu}_3[\bar{\psi}(p+k)\psi(k)h(p)]{\cal D}_{\mu\nu,\rho\sigma}(p)~,\nn\\
&=&\f{\ka^2 e}{128}\int_p(-2)\f{4\zeta p^2\gamma_\la-2(2\zeta-6)p_\la\not{p}+(6\zeta-6)\not{p}\gamma_\la\not{p}}{(p^2+i\epsilon)[(p+k)^2-m^2+i\epsilon][(p+k^\pr)^2-m^2+i\epsilon]}~,\nn\\
&=&-\f{\ka^2 e}{64}\int_p\f{(6-2\zeta) p^2\gamma_\la+8\zeta p_\la\not{p}}{(p^2+i\epsilon)[(p+k)^2-m^2+i\epsilon][(p+k^\pr)^2-m^2+i\epsilon]}~,\nn\\
&=&-i\f{e \ka^2\gamma^\la}{32}(3+\zeta)I_2~,
\eeqa
\nz
\item Diagram (2):\small
\beqa
-i\Gamma_e^{2}&=&\f{1}{2}\int_p F^{\mu\nu,\rho\si;\la}_5 {\cal D}_{\mu\nu,\rho\sigma}(p)~,\nn\\
&=&\f{\ka^2 e  }{2}\f{1}{2}\int_p\f{20(\zeta-1)p^\la \not{p}-(26\zeta+22)p^2\ga^\la}{8(p^2+i\epsilon)}~,\nn\\
&=&{ie\ka^2\gamma^\la }\f{-\zeta-2}{2}I_2,
\eeqa\nz
\item Diagram (3):\small
\beqa
-i\Gamma_e^{(3)}&=&\int_p E_3^{\mu\nu,\la}[\bar{\psi}\psi A h]\f{i(\not{p}+\not{k}+m)}{(p+k)^2-m^2+i\epsilon}D^{\rho\sigma}_3[\bar{\psi}(p+k)\psi(k)h(p)]{\cal D}_{\mu\nu,\rho\sigma}(k)~,\nn\\
&=&-\f{\ka^2}{64}\int_p\f{2[(4-8\zeta)p^\la \not{p}+(2\zeta-10)p^2\gamma^\la]}{p^2(p^2+i\epsilon)[(p+k)^2-m^2+i\epsilon]}~,\nn\\
&=&-\f{\ka^2}{32}\gamma^\la\int_p\f{[(2-4\zeta)+(2\zeta-10)]p^2}{p^2(p^2+i\epsilon)[(p+k)^2-m^2+i\epsilon]}~,\nn\\
&=&i\f{e\ka^2\gamma^\la}{16}(\zeta+4)I_2~,
\eeqa\nz
\item Diagram (4):\small
\beqa
-i\Gamma_e^{(4)}&=&\int_p D^{\rho\sigma}_3[\bar{\psi}(p+k)\psi(k)h(p)]\f{i(\not{p}+\not{k}+m)}{(p+k)^2-m^2+i\epsilon}E_3^{\mu\nu,\la}[\bar{\psi}\psi A h]{\cal D}_{\mu\nu,\rho\sigma}(k)~,\nn\\
&=&-\f{\ka^2}{64}\int_p\f{-2[(8\zeta-4)p^\la \not{p}+(10-2\zeta)p^2\gamma^\la]}{p^2(p^2+i\epsilon)[(p+k)^2-m^2+i\epsilon]}~,\nn\\
&=&-\f{\ka^2}{32}\gamma^\la\int_p\f{[(2-4\zeta)+(2\zeta-10)]p^2}{p^2(p^2+i\epsilon)[(p+k)^2-m^2+i\epsilon]}~,\nn\\
&=&i\f{e\ka^2 \gamma^\la}{16}(\zeta+4)I_2~,
\eeqa\nz
\item Diagram (5),(6),(7) give no contributions to gauge vertex corrections because they are logarithm divergent.
\eit

So the sum of the gauge vertex corrections are
\beqa
-i\Gamma&=&-i\sum\limits_{i}\Gamma_e^i~,\nn\\
&=&-ie \gamma^\la \f{\ka^2}{32}(13\zeta+19)I_2~.
\eeqa
Then the gauge vertex counter term $\delta_1$ is given by
\beqa
\delta_1=-\f{\ka^2}{32}(13\zeta+19)I_2.
\eeqa
Therefore, it is obvious from our calculation that the Ward-Takahashi identity $\delta_1=\delta_2$ is satisfied with the previous choice of regularization procedure. This check is fairly non-trivial and it is justify the use of the new regularizaiton scheme. We also check that the Ward-Takahashi identity will be spoiled if we use the DR replacement $k^\mu k^\nu\ra k^2 \eta^{\mu\nu}/4 $ for qudratic divergence terms.

\subsection{Yukawa Vertex Corrections and Scalar Self Energy Corrections}
We still need the gravitational corrections to yukawa vertex. Relevant diagrams that contribute to the yukawa vertex are listed in the right panel of fig.\ref{fig2}. The results are calculated to be:

\bit
\item  Diagram (1)\small
 \beqa
-i \Gamma_{(1)}&=&\int_k D^{\mu\nu}_3[\bar{\psi}(p_2)\psi(p_2+k)h(-k)]\f{i}{\not{p}_2+\not{k}-m+i\epsilon}(-i Y)\f{i}{\not{p}_1+\not{k}-m+i\epsilon} \nn\\
&&~~~D^{\rho\si}_3[\bar{\psi}(p_1+k)\psi(p_1)h(k)]\DD_{\mu\nu,\si\rh}(k)~,\nn\\
&=&\f{\ka^2 Y}{64}\int_k \f{4\zeta(k^2)^2-4\zeta \not{k} \not{k} \not{k} \not{k}+3(\zeta-1)k^2 \not{k} \not{k}}{[(p_1+k)^2-m^2+i\epsilon][(p_2+k)^2-m^2+i\epsilon](k^2+i\epsilon)}\nn\\
&=&\f{\ka^2 Y}{128}\int_k \f{12(\zeta-1)(k^2)^2}{[(p_1+k)^2-m^2+i\epsilon][(p_2+k)^2-m^2+i\epsilon](k^2+i\epsilon)}\nn\\
&=&i{\ka^2 Y}\f{3}{32}(\zeta-1)I_2~,
\eeqa\nz
\item Diagram (2)\sz
\beqa
-i \Gamma_{(2)}&=&\f{1}{2}\int_k D_5^{\mu\nu,\rho\sigma}\DD_{\mu\nu,\si\rh}(k)~,~~~~~~~~~~~~~~~~~~~~~~~~~~~~~~~~~~~~~~~~~~~~~~~~~~~~~~~~~\nn\\
&=&-\f{\ka^2}{8} Y\int_k \f{8(3+2\zeta)}{k^2+i\epsilon}~,\nn\\
&=&-i\ka^2 Y\f{3+2\zeta}{2}I_2~,
\eeqa\nz
\item Diagram (3)\sz
\beqa
-i \Gamma_{(3)}&=&\int_k\f{-i \ka Y \eta_{\mu\nu}}{2} \f{i}{\not{p}_1+\not{k}-m+i\epsilon} D^{\mu\nu}_3[\bar{\psi}(p_1+k)\psi(p_1)h(-k)]\DD_{\mu\nu,\si\rh}(k)~,\nn\\
&=&\f{\ka^2 Y}{16}\int_k \f{(2d^2-6d+4)\not{k}\not{k}}{[(p_1+k)^2-m^2+i\epsilon](k^2+i\epsilon)}~,\nn\\
&=&i\f{3}{8}\ka^2Y I_2~,
\eeqa\nz
\item Diagram (4)\sz
\beqa
-i \Gamma_e&=&\int_kD^{\mu\nu}_3[\bar{\psi}(p_2)\psi(p_2+k)h(-k)] \f{i}{\not{p}_2+\not{k}-m+i\epsilon}\f{-i \ka Y \eta_{\mu\nu}}{2} \DD_{\mu\nu,\si\rh}(k)~,\nn\\
&=&\f{\ka^2 Y}{16}\int_k \f{(2d^2-6d+4)\not{k}\not{k}}{[(p_2+k)^2-m^2+i\epsilon](k^2+i\epsilon)}~,\nn\\
&=&i\f{3}{8}\ka^2Y I_2~,
\eeqa
\nz
\item  Diagram (5)(6)(7) gives null contribution to the vertex corrections because of their logarithm divergence.
\eit

So the sum of vertex correction for yukawa couplings
\beqa
-i \Gamma_{\rm total}&=&-i\sum\limits_{i=1}^7 \Gamma_i~,\nn\\
&=&-\f{i\ka^2 Y}{32}\left(27+29\zeta\right)I_2~.
\eeqa

Then we get the renormalization constant
\beqa
\delta_Y=(Z_Y-1)=-\f{\ka^2 }{32}\left(27+29\zeta\right)I_2~,
\eeqa

\begin{figure}[bthp]
\label{fig3}
\begin{center}
\begin{picture}(290,90)(0,0)
\Vertex(80,40){2}
\DashLine(30,40)(80,40){3}\DashLine(80,40)(130,40){3}
\ArrowArc(80,60)(19,0,360)\ArrowArc(80,60)(21,0,360)
\Text(55,38)[t]{$p$}\Text(105,38)[t]{$p$}
\Text(80,80)[t]{$k$}
\Text(80,15)[l]{a}
\Vertex(200,40){2} \Vertex(240,40){2}
\DashLine(160,40)(200,40){2}\Text(180,38)[t]{$p$}
\DashLine(200,40)(240,40){2}\Text(220,38)[t]{$p+k$}
\DashLine(240,40)(280,40){2}\Text(260,38)[t]{$p$}
\ArrowArc(220,40)(19,0,180)\Text(220,65)[b]{$k$}
\ArrowArc(220,40)(21,0,180)
\Text(220,15)[l]{b}
\end{picture}
\end{center}
\caption{The gravitational contributions to scalar self energy.}
\end{figure}
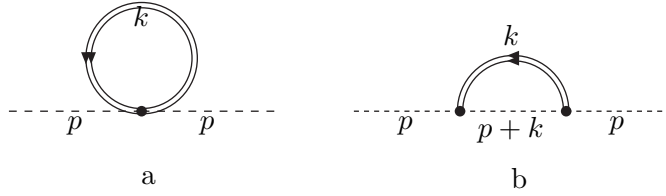

The scalar self-energy corrections with the diagrams in fig.(\ref{fig3}) had already been calculated in our previous work\cite{previous}. However, the regularization procedure used in that calculation is the ordinary gauge non-preserving cut off scheme. So we need to reformulate the previous results.

The scalar self-energy renormalization are calculated to be
 \beqa
 i\Pi(q^2)[a] &=&
\f{1}{2} \int_p C_4^{\mu\nu,\si\rh}[\phi(q)\phi(-q) hh]
 \DD_{\mu\nu,\si\rh}(p)
 \nn\\
 &=& -\f{\ka^2}{16}\int_p \f{ 8d(\zeta-1) (p\cdot q)^2-(4d\zeta-8\zeta+3d^2-18d+16)p^2 q^2}{(p^2+i\epsilon)^2}\,\nn \\
&=&-\f{\ka^2}{8}q^2\int_p  p^2\f{ 4d(\zeta-1) -(4d\zeta-8\zeta+3d^2-18d+16) }{(p^2+i\epsilon)^2}\,\nn\\
&=&-i\ka^2 q^2\f{\zeta-1}{2}I_2~,
 \label{eq:SF-a}
 \eeqa
and
\beqa
 i\Pi(q^2)[b] &=&
 \int_p
 C_3^{\mu\nu}[\phi(q)\phi(-p-q)h]
 C_3^{\si\rh}[\phi(p+q)\phi(-q)h]
 \DD_{\mu\nu,\si\rh}(p)  \nn\\
 &=&\int_p\f{\ka^2}{8}\f{4\ze q^2 p^2}{p^2+i\epsilon} \nn\\
 &=&i{\ka^2}q^2\f{\ze}{2}  I_2  \,.
 \label{eq:SF-b}
 \eeqa
which gives the counter term for scalar self-energy
\beqa
\delta_\phi\equiv Z_\phi-1=-\f{\ka^2}{2} I_2~,
\eeqa

This result is different to our previous result\cite{previous} because we use the gauge preserving cut-off regularization scheme.

With all the corrections at hand, we can obtain the RGE for yukawa couplings $Y=Y_0 Z_\psi Z_\phi^{1/2}Z_Y^{-1}$
\beqa
\beta_Y&=&\mu\f{d}{d\mu} Y(Y_0,\f{\Lambda}{\mu})=-\Lambda\f{d}{d \Lambda}Y(Y_0,\f{\Lambda}{\mu})~,\nn\\
&=&-Y \[\Lambda\f{d}{d\Lambda}\(\delta_\psi+\f{1}{2}\delta_\phi-\delta_Y\)\]~, \nn\\
&=&Y \f{\zeta}{2} \f{\ka^2\Lambda^2}{8\pi^2}~,
\eeqa
The final result is gauge dependent which is also anticipated in \cite{yukawaGD}.

If we choose the harmonic gauge fixing $\zeta=1$, we can obtain the gravitational contributions to yukawa beta function $\Delta \beta_Y=1$.
  In standard model, the yukawa coupling has the form
  \beqa
 {\cal L}\supset -\f{y_i}{\sqrt{2}} \bar{f}_L f_R h+h.c. ~,
  \eeqa
so we need the replacement $Y\ra y_i/\sqrt{2}$ for SM yukawa couplings. The beta function from gravitational contribution will not be changed with such replacement.
\subsection{Gravitational Corrections To Gauge And Scalar Couplings}
 In order to obtain the RGE for SM, we must also include the gravitational contributions to gauge and scalar couplings.

\begin{figure}[bthp]
\label{fig4}
\begin{center}
\begin{picture}(290,90)(0,0)
\Vertex(80,40){2}
\Photon(30,40)(80,40){2}{5}\Photon(80,40)(130,40){2}{5}
\ArrowArc(80,60)(19,0,360)\ArrowArc(80,60)(21,0,360)
\Text(55,38)[t]{$k$}\Text(105,38)[t]{$k$}
\Text(80,78)[t]{$p$}
\Text(80,15)[l]{a}
\Vertex(200,40){2} \Vertex(240,40){2}
\Photon(160,40)(200,40){2}{5}\Text(180,38)[t]{$k$}
\Photon(200,40)(240,40){2}{5}\Text(220,38)[t]{$p+k$}
\Photon(240,40)(280,40){2}{5}\Text(260,38)[t]{$k$}
\ArrowArc(220,40)(19,0,180)\Text(220,65)[b]{$p$}
\ArrowArc(220,40)(21,0,180)
\Text(220,15)[l]{b}
\end{picture}
\end{center}
\caption{The gravitational contributions to photon self energy.}
\end{figure}
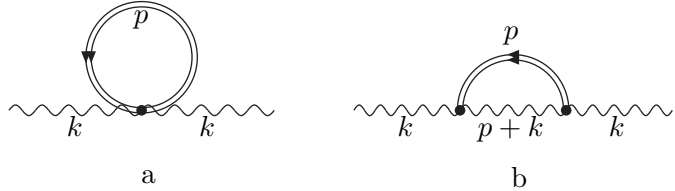
 The gravitational contributions to gauge coupling are given by the photon self-energy diagrams in fig.\ref{fig4} and are calculated to be
\beqa
i\Pi^{\mu\nu}_1&=&\f{1}{2}\int_p G^{\mu\nu,\rho\sigma;\al\beta}_4[A_\al(k) A_\bt(k)] {\cal D}_{\mu\nu,\rho\sigma}(p)~,\nn\\
&=&-3\ka^2\int_p\f{[(\zeta-1)k.p(k_\mu p_\nu+k_\nu p_\mu)+k_\mu k_\nu p^2-(\zeta-1)(k\cdot p)^2g_{\mu\nu}-k^2(\zeta-1)p_\mu p_\nu-p^2 k^2 g_{\mu\nu}]}{2p^2(p^2+i\epsilon)}~,\nn\\
&=&3\ka^2 \zeta(k^2g_{\mu\nu}-k_\mu k_\nu)\int_p \f{1}{2(p^2+i\epsilon)}~,\nn\\
&=& \f{3i\ka^2}{2} \zeta(k^2g_{\mu\nu}-k_\mu k_\nu) I_2~,
\eeqa
and
\beqa
i\Pi^{\al\la}_2&=&\int_p G^{\mu\nu;\al\beta}_3[A_\al(k) A_\bt(p+k)] G^{\rho\sigma;\la\tau}_3[A_\la(k) A_\tau(p+k)]{\cal D}_{\mu\nu,\rho\sigma}(p)\f{-i g_{\bt\tau}}{(p+k)^2+i\epsilon}~,\nn\\
&=&\ka^2\int_p\f{2\zeta k_\al k_\la p^2- k^2[2\zeta p^2 g_{\al\la} -(\zeta-3)p_\al p_\la]-(k\cdot p)(\zeta-3)( k_\la p_\al+p_\al k_\la)+(\zeta-3)(k\cdot p)^2 g_{\la\al}}{2[(p+k)^2+i\epsilon][p^2+i\epsilon]}~,\nn\\
&=&\ka^2[({3+\zeta}) k_{\al}k_{\la}-({3+\zeta})k^2g_{\al\la}]\int_p \f{p^2}{2[(p+k)^2+i\epsilon][p^2+i\epsilon]}~,\nn\\
&=&i\ka^2\f{3+\zeta}{2}(k_{\al}k_{\la}-k^2g_{\al\la})I_2~,
\eeqa

So we can obtain
\beqa
i\Pi^{\mu\nu}=i\f{2\zeta-3}{2}(k^2g_{\mu\nu}-k_{\mu}k_\nu)I_2~,
\eeqa
For harmonic gauge fixing $\zeta=1$, we can see that
\beqa
\delta_A=Z_A-1=\ka^2\f{2\zeta-3}{2} I_2=-\ka^2 I_2~,
\eeqa
Therefor the gauge coupling RGE is given by
\beqa
\beta_{A_\mu}=\mu\f{d}{d\mu}e(e_0,\f{\Lambda}{\mu})&=&-\Lambda\f{d}{d\Lambda}(\f{1}{2}\delta_A e_0)~,\nn\\
&=& -\f{e}{32\pi^2}\ka^2 \Lambda^2~,
\eeqa
Our result agree with the result given in \cite{ylwu,tom3} and also agrees with the traditional BFM (background field method) in the harmonic gauge with LORE regularization scheme\cite{ylwu}. Note that if we use $k^\mu k^\nu\ra  g_{\mu\nu} k^2/4$ for quadratic divergence terms, vanishing result will be obtained which was found in \cite{gaugedependent1}. However, such replacement of DR regularization scheme will spoil the Ward-Takahashi identity related to gauge invariance.

 We also need the beta function for scalar quartic coupling with the new regularization rule $k^\mu k^\nu\ra  g_{\mu\nu} k^2/2$  for quadratic divergence terms.
The relevant Feynman diagrams can be seen in our previous paper\cite{previous}.
The non-vanishing vertex correction is given by
\beqa
 i\Gamma_4 &=& \f{1}{2}\f{i\ka^2\la}{4}
 \(\et^{\mu\si}\et^{\nu\rh}+
   \et^{\mu\rh}\et^{\nu\si}-
   \et^{\mu\nu}\et^{\si\rh}\)
   \int_p \DD_{\mu\nu,\si\rh}(p)
 \nn\\[2mm]
 &=& -\f{\ka^2\la}{16}8(3+2\zeta)\int_p \f{1}{p^2+i\epsilon} \,,\nn\\
 &=&-{\ka^2\la}\f{3+2\ze}{2} i I_2\,
 \eeqa
The counter term $\delta_\la$ is given by
\beqa
\delta_\la=-\ka^2\la\f{3+2\zeta}{2} I_2~.
\eeqa
So we can obtain the beta function for quartic coupling
\beqa
\beta_\la=\mu\f{d}{d\mu}\la(\la_0,\f{\Lambda}{\mu})&=&-\Lambda\f{d}{d \Lambda}(\la_0 Z_\phi^2-\delta_\la)~,\nn\\
&=&\Lambda\f{d}{d \Lambda}\f{1+2\zeta}{2}\f{\la_0\ka^2\Lambda^2}{16\pi^2}~,\nn\\
&=&(1+2\zeta)\la_0\f{\ka^2\Lambda^2}{16\pi^2}~.
\eeqa
In harmonic gauge with $\zeta=1$, we can obtain
\beqa
\beta_\la\equiv\f{d\la}{d\ln E}=\f{3}{16\pi^2}\la \ka^2 E^2~,
\eeqa

 We should note that the results depend on the gravitational gauge fixing choice. In certain circumstances, the Vilkovisky-DeWitt effective action approach which is intrinsically gauge-preserving can agree with the traditional approach in harmonic gauge\cite{ylwu}. So it is preferable to fix the physical result in harmonic gauge with $\zeta=1$ and such results are used in our subsequent calculations.

\section{ MPCP Constraints With Gravitational Contributions}
Constraints from MPCP can be fairly predictive and they may reveal the existence of asymptotic safety of gravity. Knowing the RGE running of SM couplings, the requirement of MPCP near the Planck scale can be studied. On the other hand, the presence of new terms in the SM beta functions from gravitational effects can have important consequences. Such term can be dominant near the Planck mass scale and significantly change the running behavior of SM couplings in the UV region.

In order to study the RGE running of quartic coupling $\la$ below the Planck scale, we adopt the full two-loop Standard Model beta functions\cite{twoloop} for $\la$ (three loop results can be seen in\cite{threeloop1,threeloop2}), the top-yukawa couplings $y_t$ and gauge couplings $g_i (i=1,2,3)$  in addition to the one-loop power-law-running gravitational contributions. For the weak scale input, the following boundary conditions\cite{pdg} for RGE running are used
\beqa
\al_s(M_Z)&=&0.1185\pm0.0006\nn\\
\al_{\rm em}^{-1}(M_Z)&=&127.906\pm0.019\nn\\
\sin^2\theta_W(M_Z)&=&0.2312\pm0.0002,\nn\\
M_{\rm higgs}&=&125.9\pm0.4{\rm  GeV}.\nn\\
y_b(M_Z)&=&0.0162834,\nn\\
y_\tau(M_t)&=&0.0102.
\eeqa
which give
\beqa
\al_2(M_Z)&=&\al_{em}(M_Z)/\sin^2\theta_W=(29.5718)^{-1},\nn\\
\al_1(M_Z)&=&\al_{em}(M_Z)/\cos^2\theta_W=(98.3341)^{-1}.
\eeqa
The two loop RGE running for gauge couplings in the SM are given by
\beqa
\f{d }{d\ln E}g_i=\f{b_i}{16\pi^2}g_i^3+\f{g_i^3}{(16\pi^2)^2}\[\sum\limits_{k}b_{ki}g_k^2-Tr\(C_k^U F_U^\da F_U+C_k^D F_D^\da F_D+C_k^L F_L^\da F_L\)\].
\eeqa
with
\beqa
b_{ki}=\(\bea{ccc}\f{199}{50}&\f{9}{10}&\f{11}{10}\\\f{27}{10}&\f{35}{6}&\f{9}{2}\\\f{44}{5}&12&-26\eea\)~,~
C_k^U=\(\f{17}{10},\f{3}{2},2\)~, C_k^D=\(\f{1}{2},\f{3}{2},2\)~,C_k^L=\(\f{3}{2},\f{1}{2},0\).
\eeqa
The standard GUT normalization $g_1^2=\f{5}{3}g_Y^2$ is used in the previous expressions. We neglect the yukawa couplings for the first two generations and keep the third generation contributions. The RGEs  for $g_i$ are given by
\beqa
\f{d g_1}{d\ln E}=\f{g_1}{16\pi^2}\(\f{41}{10}g_1^2-\f{\ka^2E^2}{2}\)+\f{g_1^3}{(16\pi^2)^2}\[\f{199}{50}g_1^2+\f{27}{10}g_2^2+\f{44}{5}g_3^2-\f{17}{10}y_t^2-\f{1}{2}y_b^2-\f{3}{2}y_\tau^2\].\nn
\eeqa
and
\beqa
\f{d g_2}{d\ln E}=-\f{g_2}{16\pi^2}\(\f{19}{6}g_2^2+\f{\ka^2E^2}{2}\)+\f{g_2^3}{(16\pi^2)^2}\[\f{9}{10}g_1^2+\f{35}{6}g_2^2+12g_3^2-\f{3}{2}y_t^2-\f{3}{2}y_b^2-\f{1}{2}y_\tau^2\].\nn
\eeqa
as well as
\beqa
\f{d g_3}{d\ln E}=-\f{g_3}{16\pi^2}\(7g_3^2+\f{\ka^2E^2}{2}\)+\f{g_3^3}{(16\pi^2)^2}\[\f{11}{10}g_1^2+\f{9}{2}g_2^2-26g_3^2-2y_t^2-2y_b^2\].\nn
\eeqa

 The two-loop RGE for top-yukawa coupling with one loop gravitational contribution is given by
\beqa
\f{d}{d\ln E} y^t&=&\f{1}{16\pi^2}\(\f{9}{2}y_t^3+\f{3}{2}y_b^2y_t+y_{\tau}^2y_t-8g_3^2y_t-\f{9}{4}g_2^2y_t-\f{17}{20}g_1^{ 2}y_t+\ka^2 E^2y_t\)\nn\\&+&\f{1}{(16\pi^2)^2}\[-12y_t^4-\f{11}{4}y_t^2y_b^2-\f{1}{4}y_b^4+\f{5}{4}y_b^2y_\tau^2-\f{9}{4}y_t^2y_\tau^2
-\f{9}{4}y_\tau^4+6\la^2-12\la y_t^2-4\la y_b^2\.\nn\\
&+&(\f{393}{80}g_1^2+\f{225}{16}g_2^2+36g_3^2)y_t^2
+\(\f{7}{80}g_1^2+\f{99}{16}g_2^2+4g_3^2\)y_b^2+\(\f{15}{8}g_1^2+\f{15}{8}g_2^2\)y_{\tau}^2\nn\\
&+&\left.\f{1187}{600}g_1^4-\f{9}{20}g_1^2g_2^2+\f{19}{15}g_1^2g_3^2-\f{23}{4}g_2^4+9g_2^2g_3^2-108g_3^4\].
\eeqa
 The yukawa couplings of bottom and tau gives very small contributions to RGE of $\la$, so we use here only the one-loop results
\beqa
\f{d}{d\ln E}y^b&=&\f{1}{16\pi^2}\(\f{9}{2}y_b^3+\f{3}{2}y_t^2y_b+y_{\tau}^2y_b-\f{1}{4}g_1^2y_b-\f{9}{4}g_2^2y_b-8g_3^2y_b\),~\nn\\
\f{d}{d\ln E}y^{\tau}&=&\f{1}{16\pi^2}\(\f{5}{2}y_{\tau}^3+3y_t^2y_\tau+3y_b^2y_\tau-\f{9}{4}g_1^2-\f{9}{4}g_2^2\).
\eeqa
The normalization of higgs potential in our study is given as
\footnote{ There is a factor 1/2 difference for $\la$ with respect to the convention used in \cite{luomx}.}
\beqa
V=-\f{m^2}{2} h^2+\f{\la}{4}h^4~.
\eeqa
The improved two-loop RGE for $\lambda$ with one loop gravitational contribution is given by
\beqa
\f{d}{d\ln E} \lambda=\f{1}{16\pi^2}\beta_\la^1+\f{1}{(16\pi^2)^2}\beta_\la^2~,
\eeqa
 within which the one loop part is
\beqa
\beta_\la^1&=&24\la^2-\(\f{9}{5}g_1^2+9g_2^2\)\la+\f{1}{2}\(\f{27}{100}g_1^4+\f{9}{10}g_1^2g_2^2+\f{9}{4}g_2^4\)+4\la(3y_t^2+3y_b^2+y_\tau^2)\nn\\
 &-&2(3y_t^4+3y_b^4+y_\tau^4)+3\la \kappa^2 E^2~,
\eeqa
and the two loop part\cite{luomx} is
\beqa
\beta_\la^2&=&-312\la^3+(\f{108}{5}g_1^2+108g_2^2)\la^2-\(\f{73}{8}g_2^4-\f{117}{20}g_1^2g_2^2-\f{1887}{200}g_1^4\)\la\nn\\
&+&\f{1}{2}\(\f{305}{8}g_2^6-\f{289}{40}g_2^4g_1^2-\f{1677}{200}g_2^2g_1^4-\f{1491}{1000}g_1^6\) \nn\\&-&32 g_3^2(y_t^4+y_b^4)-\f{4}{5}g_1^2\(2y_t^4-y_b^4+3y_\tau^4\)-\f{3}{4}g_2^4(3y_t^2+3y_d^2+y_\tau^2)\nn\\
&+&10\la\[\(\f{17}{20}g_1^2+\f{9}{4}g_2^2+8g_3^2\)y_t^2+\(\f{1}{4}g_1^2+\f{9}{4}g_2^2+8g_3^2\)y_b^2+\(\f{3}{4}g_1^2+\f{3}{4}g_2^2\)y_\tau^2\]\nn\\
&+&\f{g_1^2}{2}\[\(\f{63}{5}g_2^2-\f{171}{50}g_1^2\)y_t^2+\(\f{27}{5}g_2^2+\f{9}{10}g_1^2\)y_b^2+\(\f{33}{5}g_2^2-\f{9}{2}g_1^2\)y_\tau^2\]
-48\la^2(3y_t^2+3y_b^2+y_\tau^2)\nn\\
&-&\la(3y_t^4+3y_b^4+y_\tau^4)-42\la y_t^2y_b^2+10\(3y_t^6+3y_b^6+y_\tau^6\)-6y_t^4y_b^2-6y_t^2y_b^4.
\eeqa

 The inputs for RGE, namely the $\overline{\rm MS}$ gauge coupling, higgs self coupling and top yukawa coupling at the pole top mass, can be obtained with \cite{higgsbound}
\beqa
\la(m_t)&=&0.12577 + 0.00205\(\f{m_{higgs}}{\rm GeV}-125\)-0.00004\(\f{m_t}{\rm GeV}-173.15\)\pm 0.00140_{th},\nn\\
y_t(m_t)&=&0.93587 + 0.00557\(\f{m_t}{\rm GeV}-173.15\)-0.00003\(\f{m_{higgs}}{\rm GeV}-125\)\nn\\
      &-&0.00041\(\f{\al_s(M_Z)-0.1184}{0.0007}\)+0.00200_{th},\nn\\
g_3(m_t)&=& 1.1645 + 0.0031\(\f{\al_s(M_Z)-0.1184}{0.0007}\)-0.00046\(\f{m_t}{\rm GeV}-173.15\).
\eeqa
 An illustration of full gravitational contributions to the RGE running of $\la$ are shown in fig.5. We can see that the gravitational contributions greatly modify the RGE trajectory near the Planck scale.
\begin{figure}[htbp]
  \begin{minipage}[t]{0.5\linewidth}
    \centering
    \includegraphics[width=3in]{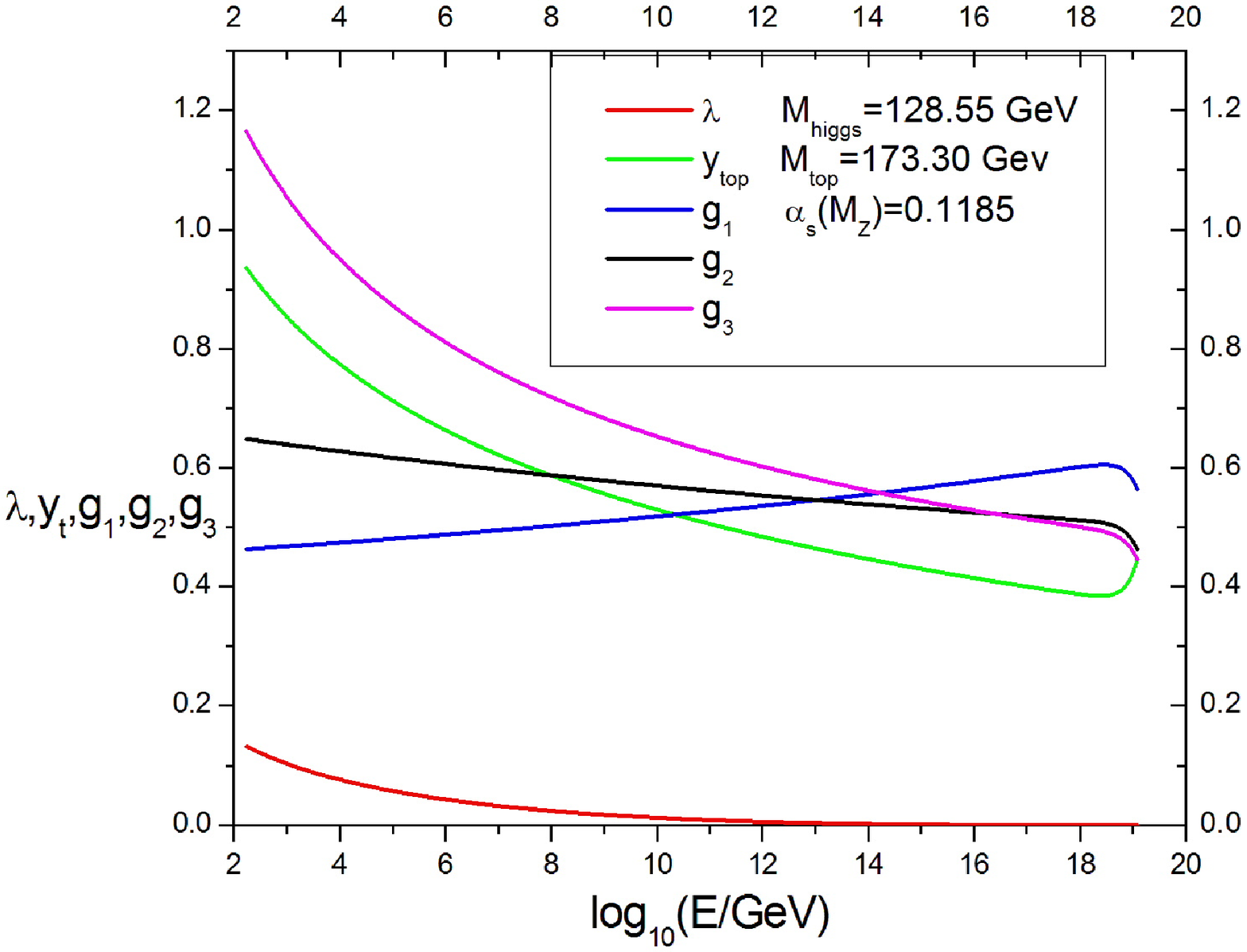}
  \end{minipage}
  \begin{minipage}[t]{0.5\linewidth}
    \centering
    \includegraphics[width=3in]{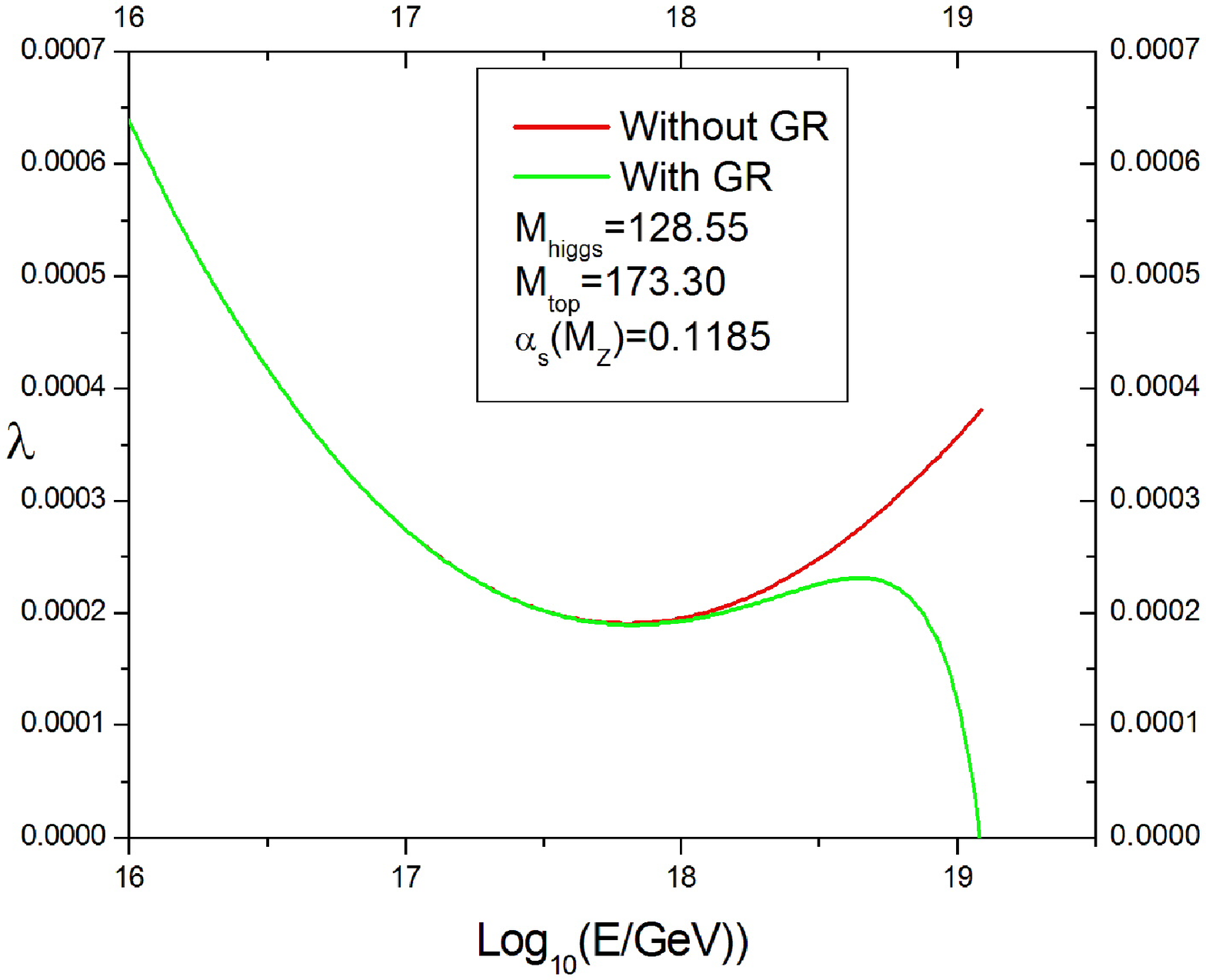}
  \end{minipage}
\caption{The left panel denotes the gravitational contributions to the RGE running of various couplings: $\la,y_t,g_1,g_2,g_3$.
The right panel compares the RGE of quartic coupling $\la$ near the Planck scale with and without gravitational couplings.}
\label{fig1}
\end{figure}

 We scan the parameter space of ($M_h,M_t$) to check the points which can lead to MPCP at some transition scale near $M_{Pl}$. We define the Planck-scale dominated region to be: $8.0\times10^{17} GeV \leq E \leq M_{Pl}$.  We use the following input:
\bit
\item  The existence of degenerated vacua near the Planck scale can be satisfied if $|\beta_\la(E)|\leq 1.0\tm 10^{-7}$, $|\la(E)|\leq 1.0\tm 10^{-4}$ and also $d^2\la(E)/d E^2\geq 0$.

\begin{figure}[htbp]
  \begin{minipage}[t]{0.3\linewidth}
    \centering
    \includegraphics[width=2 in]{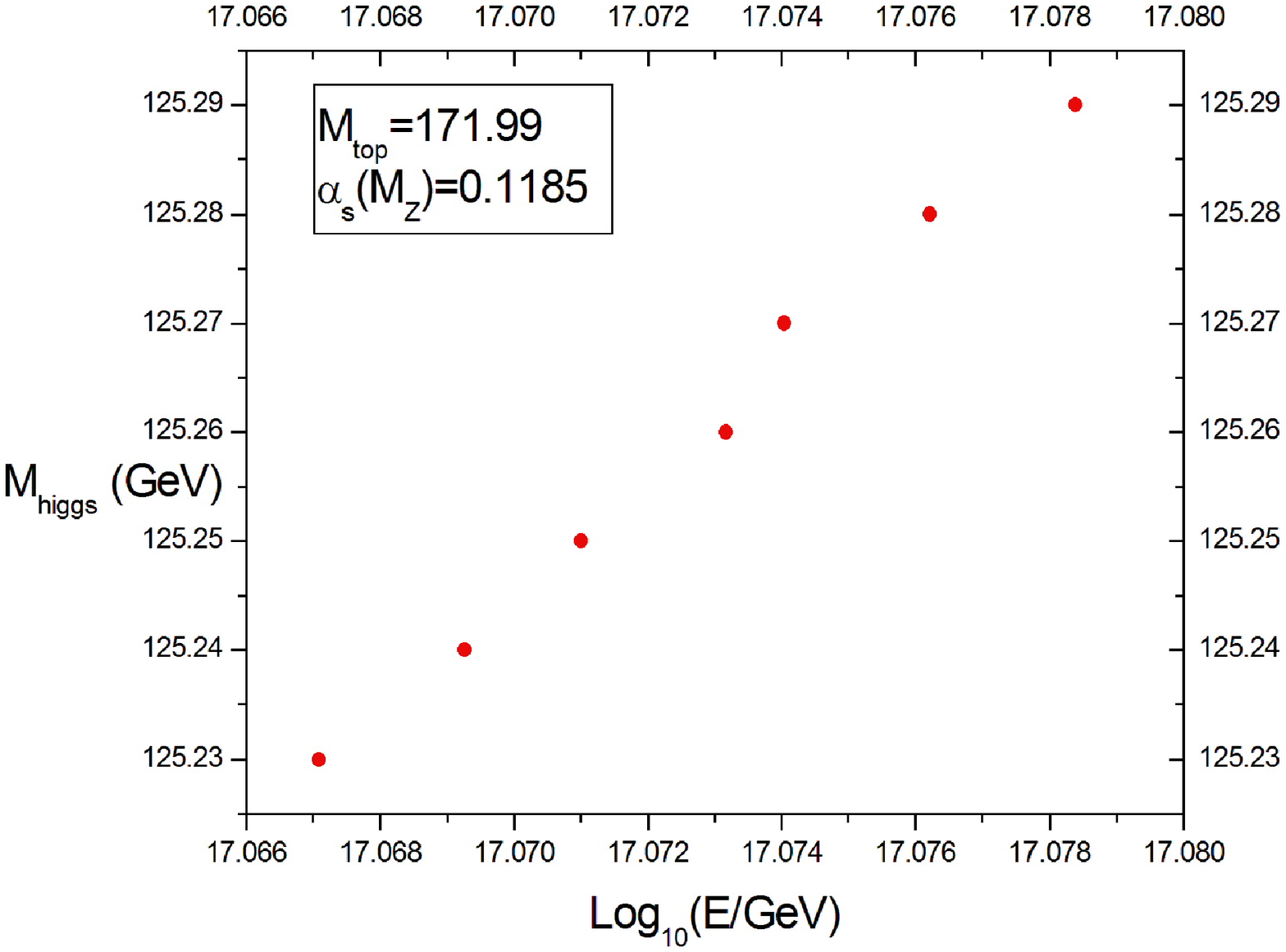}
  \end{minipage}
  \begin{minipage}[t]{0.3\linewidth}
    \centering
    \includegraphics[width=2 in]{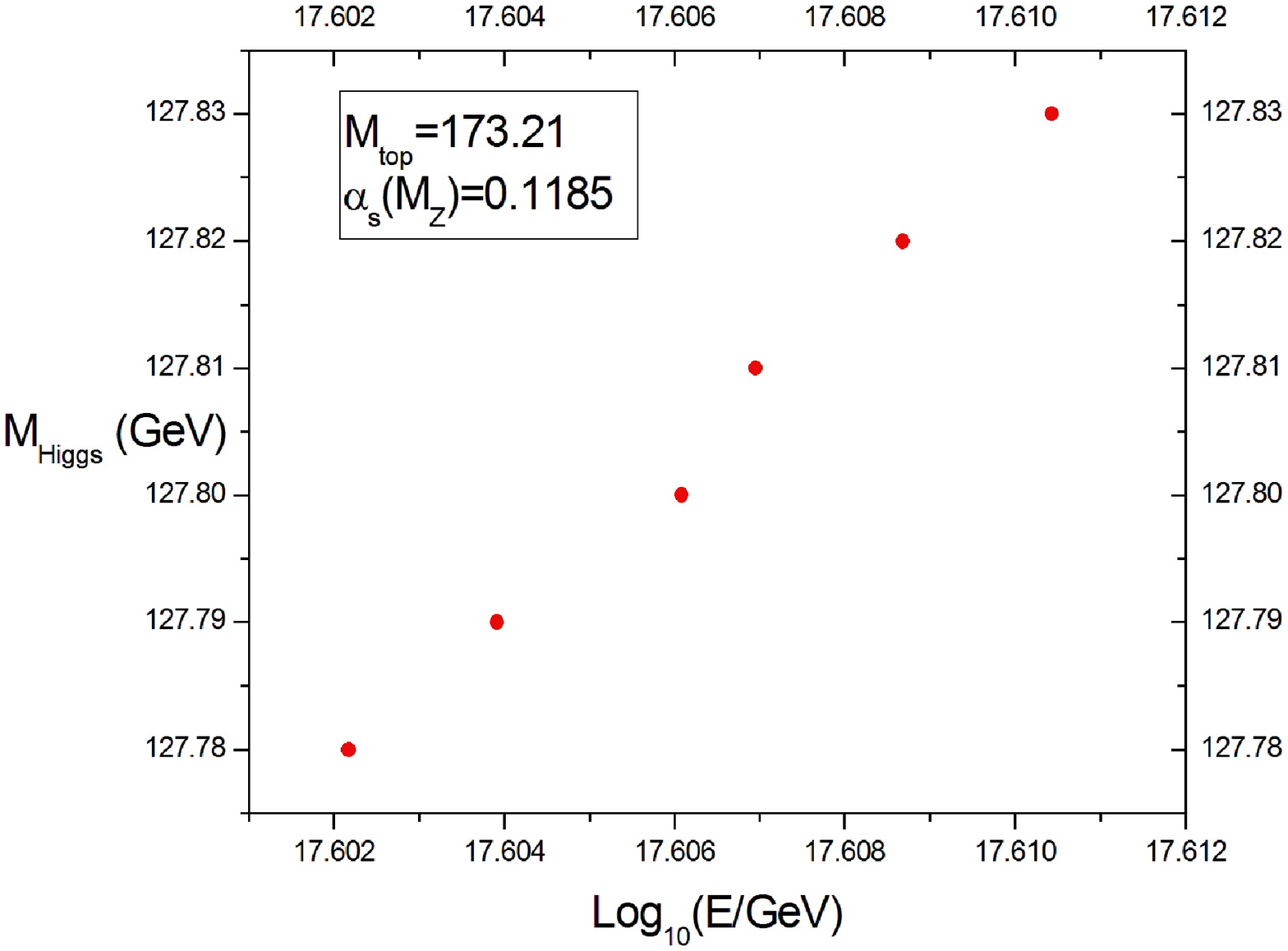}
  \end{minipage}
  \begin{minipage}[t]{0.3\linewidth}
    \centering
    \includegraphics[width=2 in]{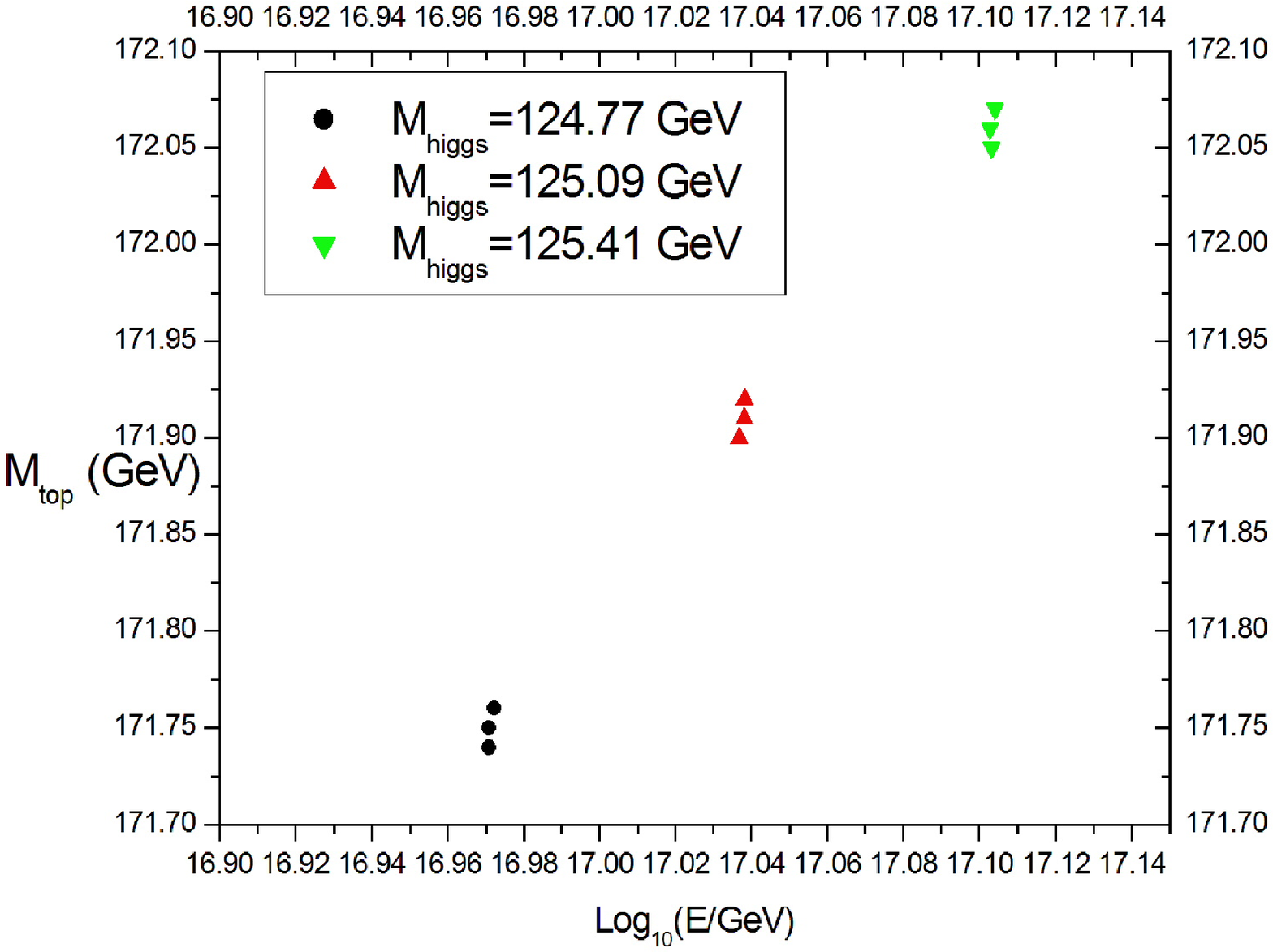}
  \end{minipage}
\caption{The points with current measured higgs and top quark masses that satisfy the MPCP conditions with the setting $|\beta_\la(E)|\leq 1.0\tm 10^{-7}$ and $|\la(E)|\leq 1.0\tm 10^{-4}$.}
\label{fig6}
\end{figure}

Our scan indicate that the MPCP condition at the Planck scale dominated region can not be satisfied with current top quark mass $m_t\in (173.21\pm 1.22) $ GeV or higgs mass $m_h\in (125.09\pm 0.32)$ GeV. The numerical results for the MPCP constraints with current measured top or higgs values are shown in fig.\ref{fig6}. We can see that MPCP could be satisfied at approximately $10^{17}-10^{17.6}$ GeV for the current measured values of top and higgs masses. Gravitational effects are negligible in finding the large-field-value degenerate vacua located in this region.

\begin{figure}[htbp]
  \begin{minipage}[t]{0.5\linewidth}
    \centering
    \includegraphics[width=3.0 in]{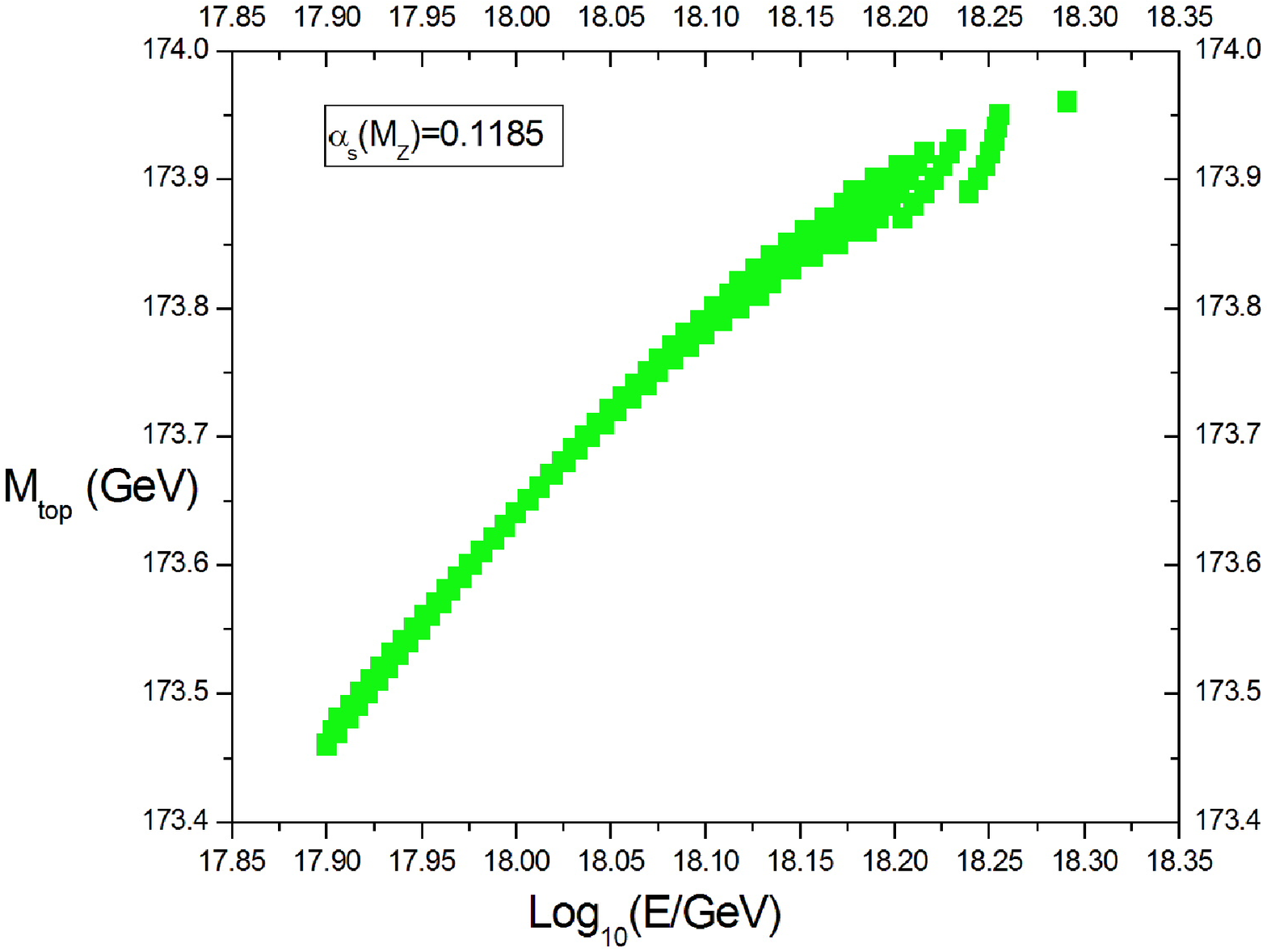}
  \end{minipage}
  \begin{minipage}[t]{0.5\linewidth}
    \centering
    \includegraphics[width=3.0 in]{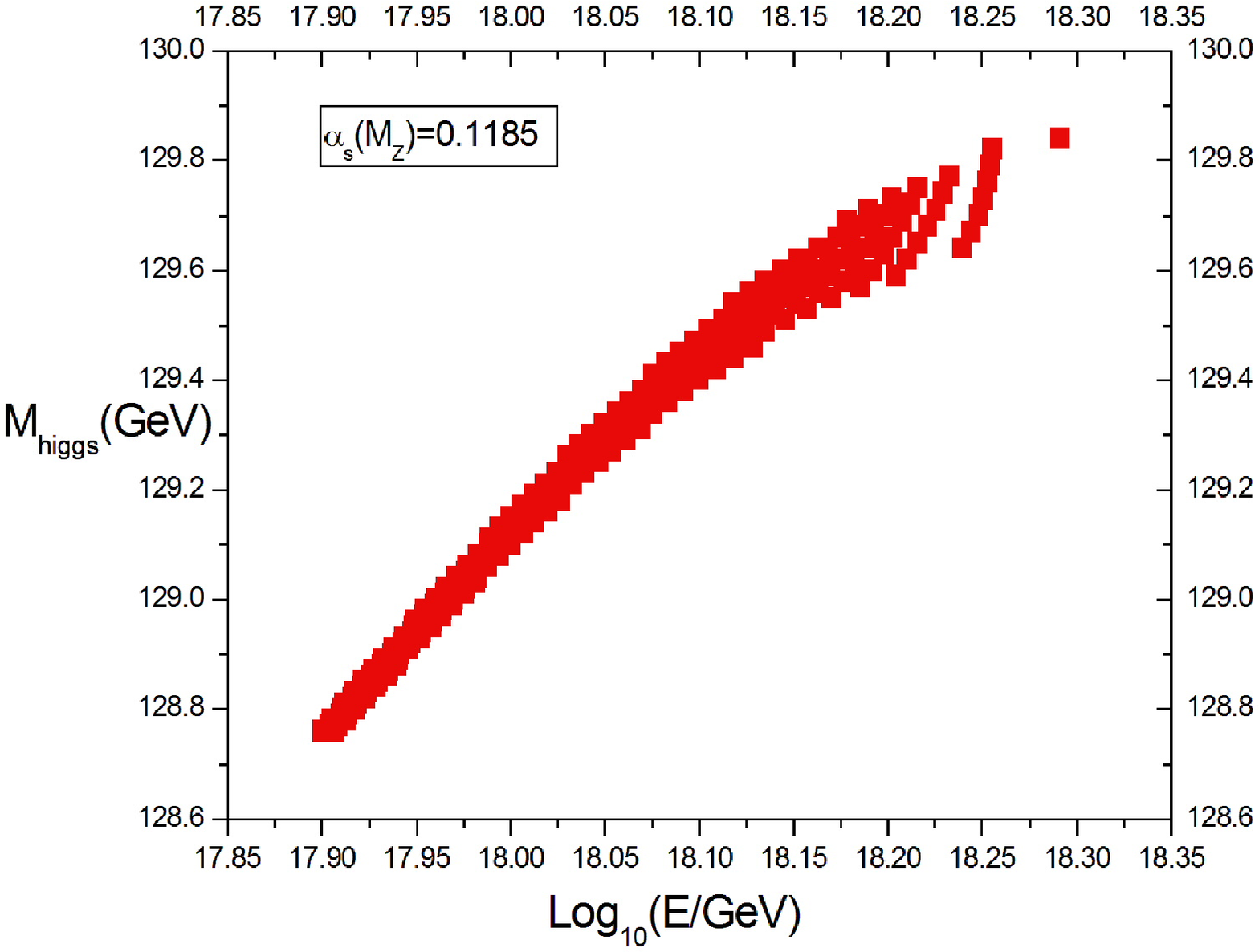}
  \end{minipage}
\caption{ The results of the scan in the parameter space $m_{higgs}\in [120,135]$ and $m_{top}\in[165,180]$ to seek the degenerated vacua in the Planck-dominated range ($8\times 10^{17}\leq E\leq M_{Pl}$). Each panel denotes the projection of the surviving triples $(E,m_t,m_h)$ onto the corresponding plane.}
\label{fig7}
\end{figure}

\item  We scan the parameter space for higgs range $m_h\in [120,135]$ GeV and top range $m_t\in[165,180]$ GeV. We keep those points that correspond to new degenerate vacua in the range $8.0\times 10^{17}\leq E\leq M_{Pl}$ (the Planck-scale dominated region). The results can be seen in the left panel of  fig.\ref{fig7}. From the the left panel, we can see that the higgs mass is constrained to lie in the range [129.0,130.2] GeV while the top quark mass in the range [173.8,174.4] GeV.
    Our scan also indicates that the highest degenerated vacua lies at approximately $2.2\tm 10^{18}$ GeV. However, if we relax the MPCP criterion to  $|\la(E)|\leq 1.0\tm 10^{-3}$, it is possible to obtain the other degenerate vacua at the Planck scale with multiple choices of top quark and higgs mass, for example, $m_t=175.10$ and  $m_h=132.14$. Constraints for MPCP in the Planck-scale dominated region also give an upper bounds on higgs and top quark masses as
$m_h\leq 134.9$ and $m_t\leq 176.34$, respectively.

\begin{figure}[htbp]
  \begin{minipage}[t]{0.3\linewidth}
    \centering
    \includegraphics[width=2 in]{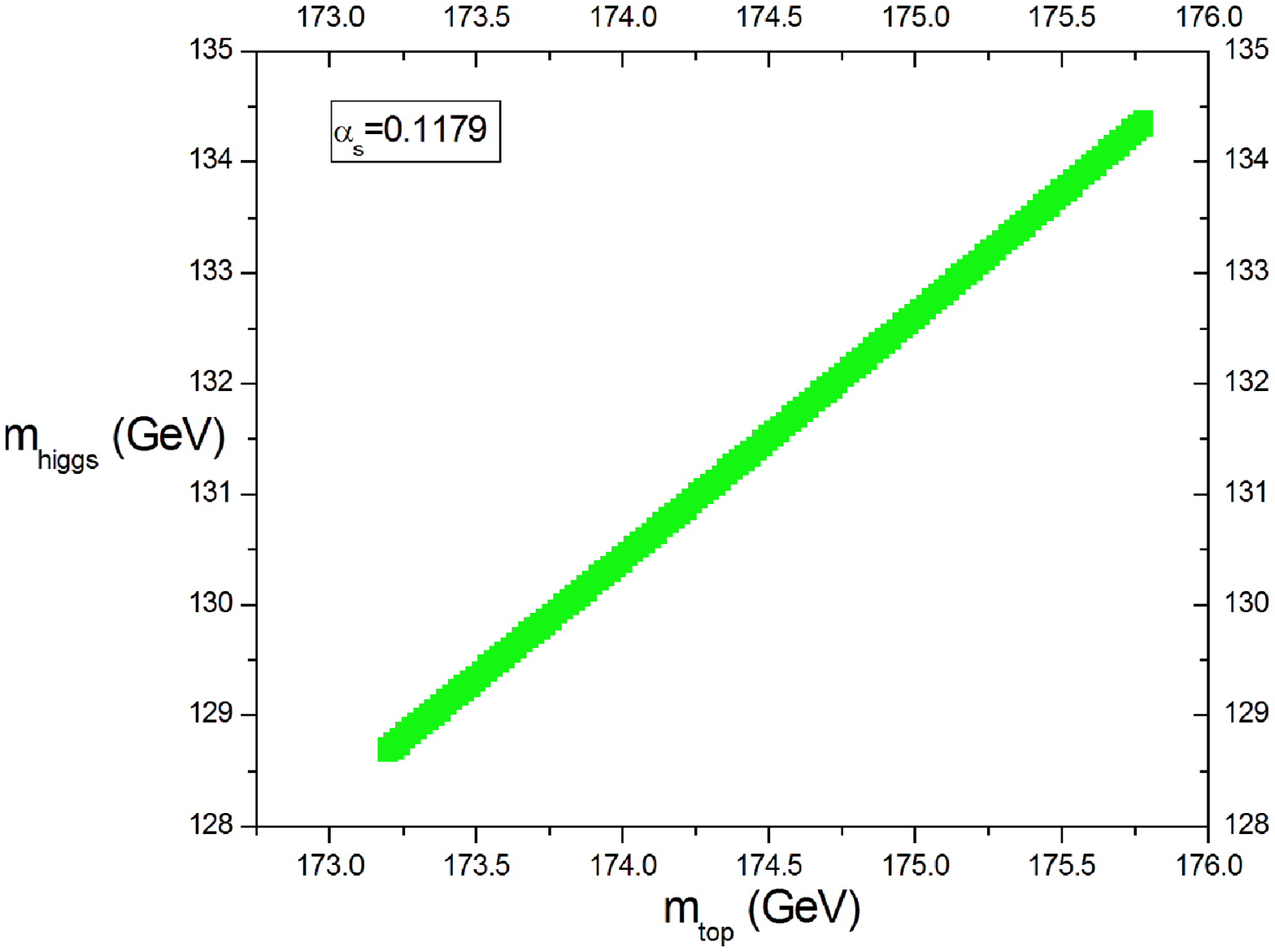}
  \end{minipage}
  \begin{minipage}[t]{0.3\linewidth}
    \centering
    \includegraphics[width=2 in]{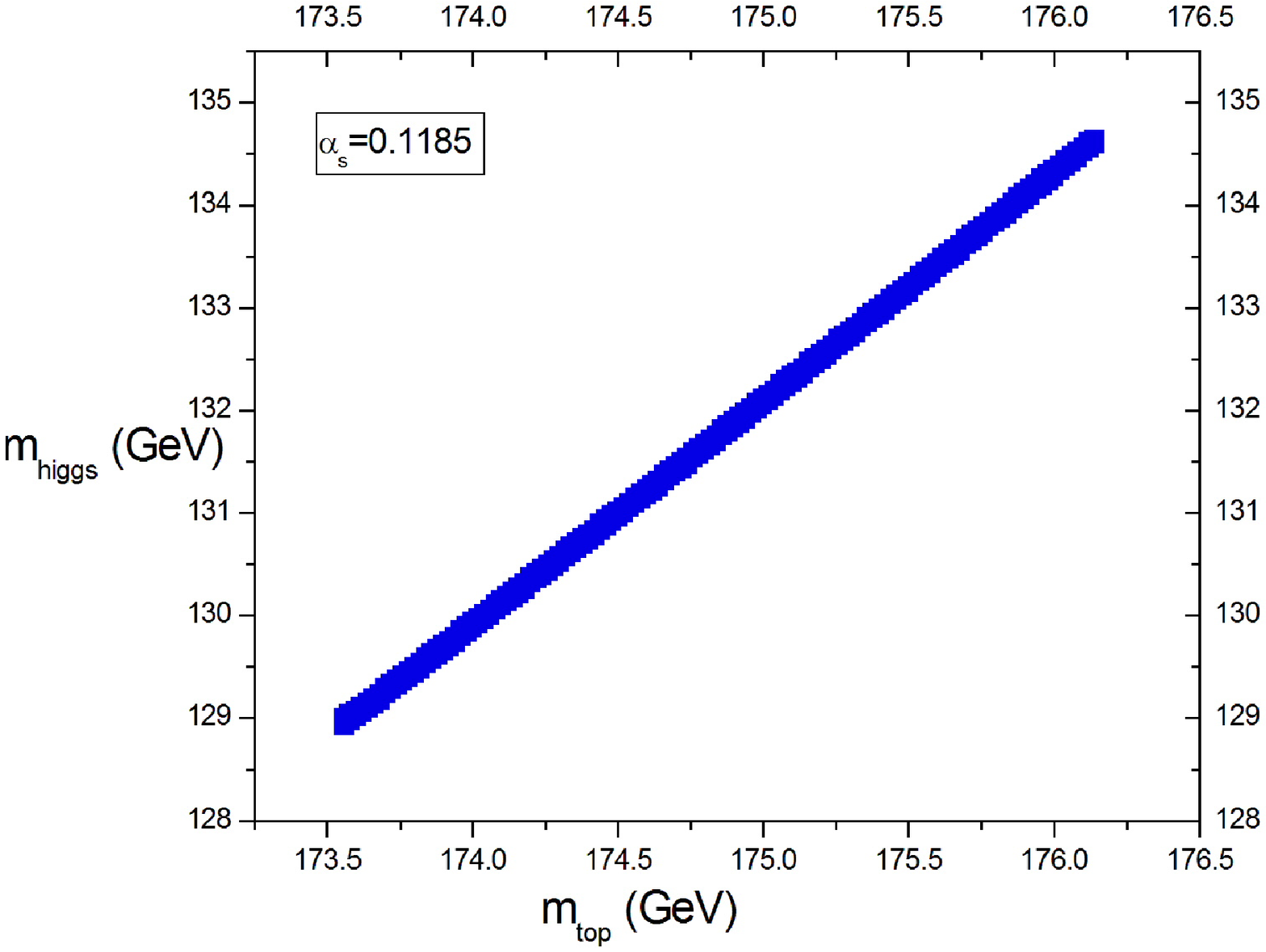}
  \end{minipage}
  \begin{minipage}[t]{0.3\linewidth}
    \centering
    \includegraphics[width=2 in]{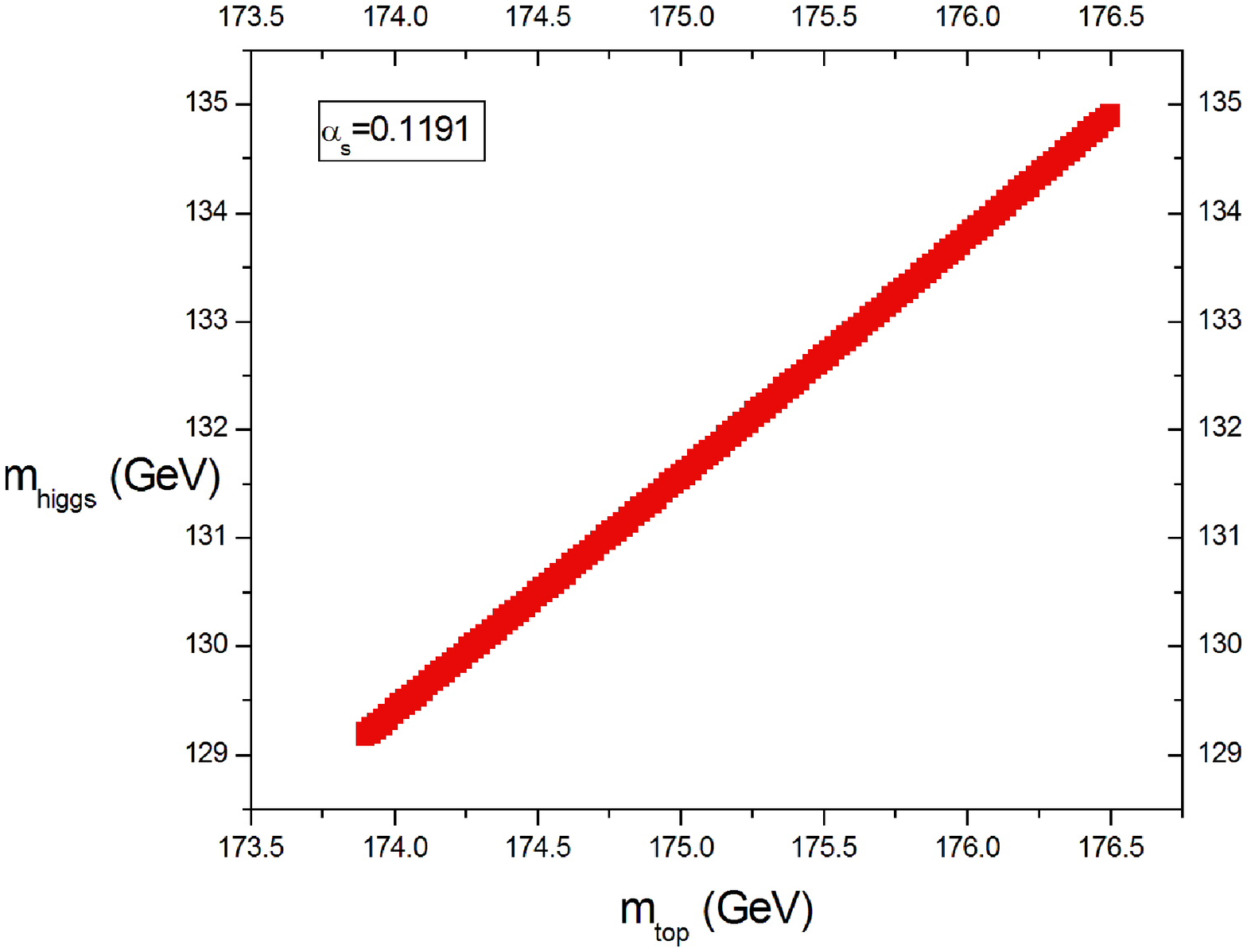}
  \end{minipage}\\
 \begin{minipage}[t]{0.3\linewidth}
    \centering
    \includegraphics[width=2 in]{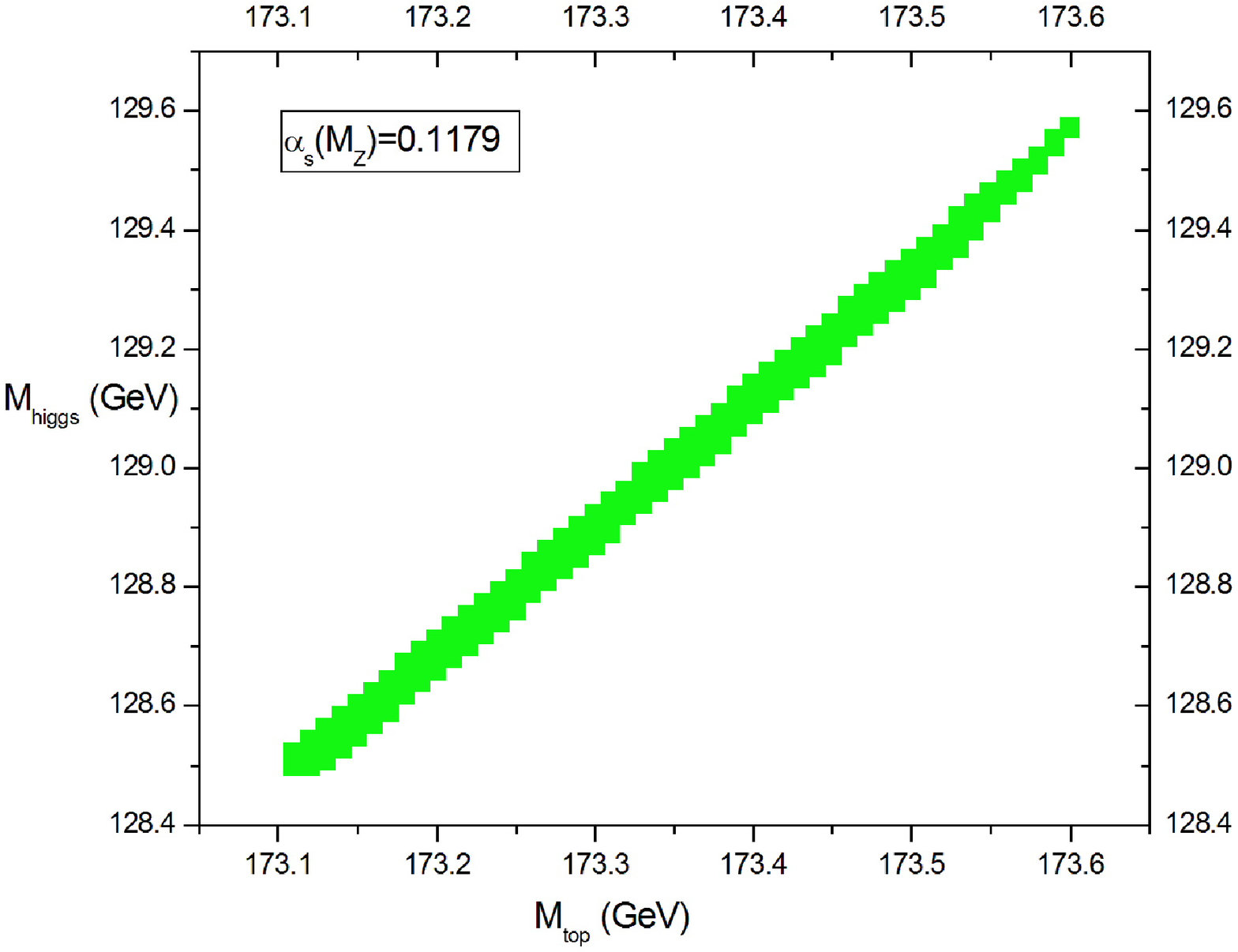}
  \end{minipage}
  \begin{minipage}[t]{0.3\linewidth}
    \centering
    \includegraphics[width=2 in]{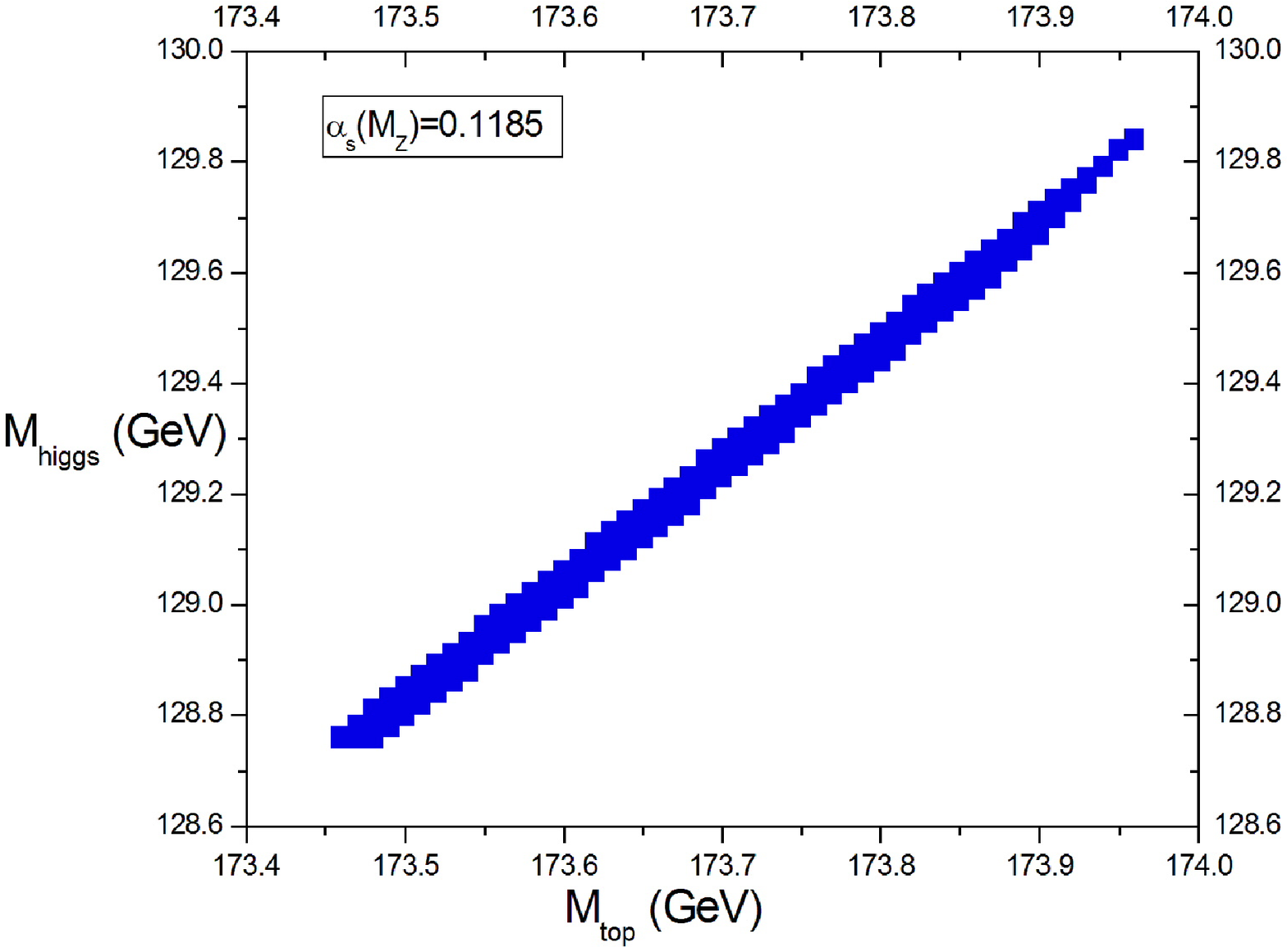}
  \end{minipage}
  \begin{minipage}[t]{0.3\linewidth}
    \centering
    \includegraphics[width=2 in]{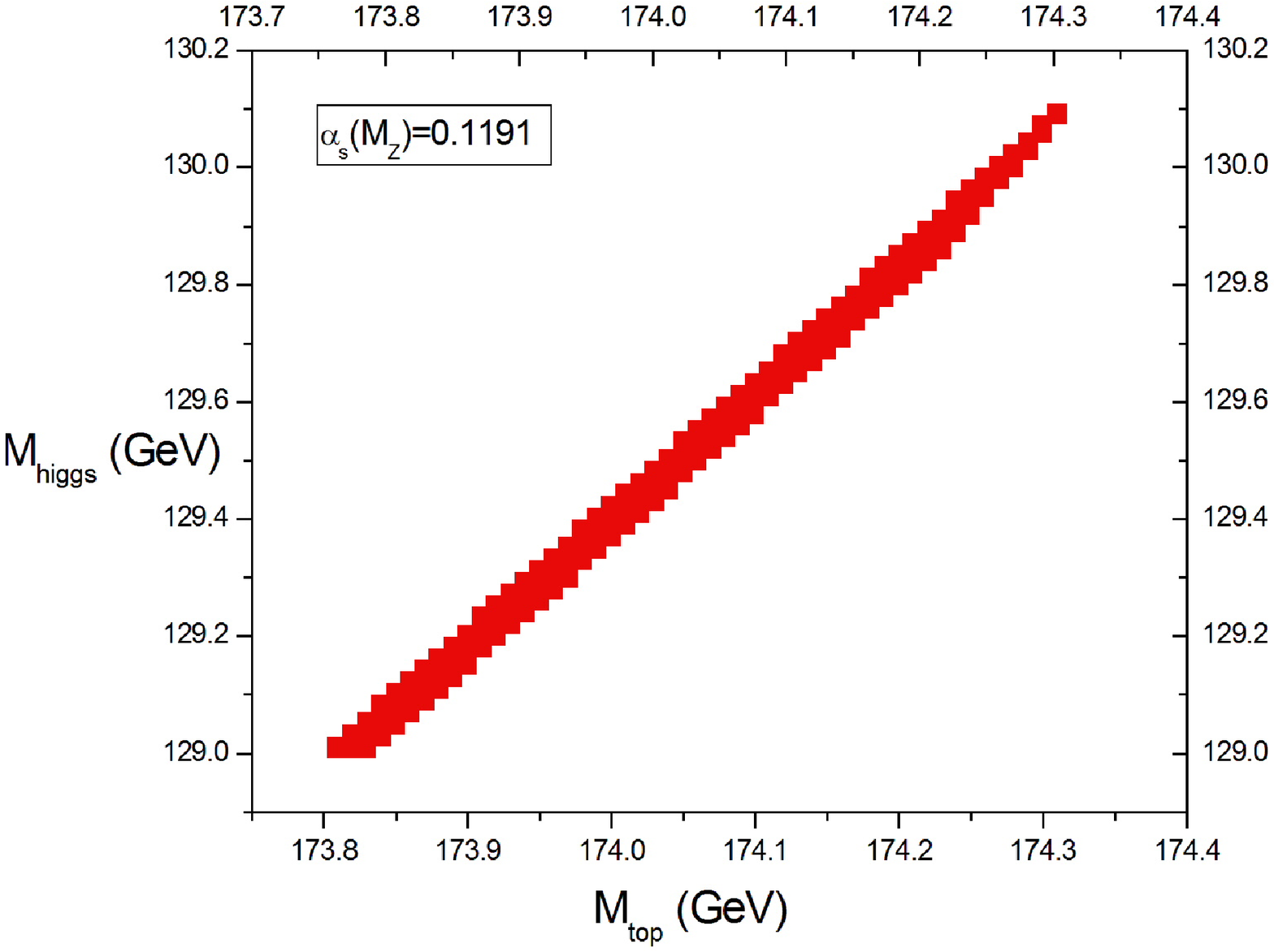}
  \end{minipage}
\caption{ Dependence on the value of $\al_s(M_Z)$ for higgs and top quark masses that can lead to the other degenerated vacua in the Planck-dominated range ($8\times 10^{17}\leq E\leq M_{Pl}$) without(the upper panel) and with (the lower panel) gravitational effects.}
\label{fig8}
\end{figure}

\item  It is instructive to compare our scan result with the case without gravitational contributions. Without gravitational effects, the bounds for higgs mass and top quark masses can be released.  We can see from fig.\ref{fig8} that the higgs mass within [129,135] GeV and top quark within [173,176] GeV can lead to degenerated vacua in the Planck-dominated region.

\item  Our scan also show that the results are sensitive to the value of $\al_s(M_Z)$. We show the results of our scan with different choice of $\al_s(M_Z)$ in fig.\ref{fig8}. Larger $\al_s(M_Z)$ requires larger $m_{top}$ and $m_{higgs}$ masses.

\eit

 \section{\label{sec-3}Conclusions}
    Based on the weak coupling expansion of gravity, we calculate the gravitational contributions to yukawa coupling, scalar quartic coupling as well as gauge couplings with general Landau-DeWitt gauge-fixing choice and a gauge preserving (of SM gauge group) cut off regularization scheme. We find that the results depend on the Landau-DeWitt gauge-fixing parameter. Based on the two loop RGE of SM couplings with one loop full gravitational contributions in harmonic gauge, we study the constraints on the higgs and top quark mass from the requirement of existing the other degenerate vacua at the Planck-dominated region. Our numerical calculations show that nature will not develop the other degenerate vacua at the Planck-dominated region with current higgs and top quark masses. On the other hand, requiring the existence of the other degenerate vacua at the Planck-dominated region will constrain the higgs and top mass to lie at approximately 130 and 174 GeV, respectively.

\begin{acknowledgments}
We acknowledge Prof. Hong-jian He and Dr. Zhong Ming for discussions and early stage cooperations. This work was supported by the Natural Science Foundation of China under grant numbers 11105124, 11105125; by the Open Project Program of State Key Laboratory of Theoretical Physics, Institute of Theoretical Physics, Chinese Academy of Sciences, P.R. China (No.Y5KF121CJ1); by the Innovation Talent project of Henan Province under grant number 15HASTIT017 and  the Young-Talent Foundation of ZhengZhou University (1421317054,1421317053).
\end{acknowledgments}

\section*{Appendix A: Feynman rules}
To simply the lengthy expressions in certain Feynman rules, we define the following expressions:
\bit
\item  $C_{\mu\nu,\rho\sigma}=\et_{\mu\rho}\et_{\nu\sigma}+\et_{\mu\sigma}\et_{\nu\rho}-\et_{\mu\nu}\et_{\rho\sigma}$.
\item $\tl{C}_{\mu\nu,\rho\sigma}(p)= p_\mu p_\rho\eta_{\nu\sigma}+p_\mu p_\sigma\eta_{\nu\rho}+p_\nu p_\rho\eta_{\mu\sigma}+p_\nu p_\sigma\eta_{\mu\rho}$\,
\item $D_{\mu\nu,\rho\sigma} (k_1, k_2)= \eta_{\mu\nu}
k_{1\sigma}k_{2\rho} - \biggl[\eta_{\mu\sigma} k_{1\nu} k_{2\rho}
  + \eta_{\mu\rho} k_{1\sigma} k_{2\nu}
  - \eta_{\rho\sigma} k_{1\mu} k_{2\nu}
  + (\mu\leftrightarrow\nu)\biggr] $
\item $C_{\mu\nu,\rho\sigma\mid\omega\tau}={\frac{1}{2}}[\eta_{\mu\omega}
C_{\rho\sigma,\nu\tau}+\eta_{\sigma\omega}C_{\mu\nu,\rho\tau}+
\eta_{\rho\omega}C_{\mu\nu,\sigma\tau}
+\eta_{\nu\omega}C_{\mu\tau,\rho\sigma} -\eta_{\omega\tau}C_{\mu\nu,\rho\sigma}+(\omega \leftrightarrow
\tau)]$
\item $ H_{\mu\nu\rho\sigma\omega\tau}(k_1,k_2)=\\
-\biggl[C_{\mu\nu,\rho\tau}
k_{1\sigma}k_{2\omega}
+C_{\mu\nu,\rho\omega}k_{1\tau}k_{2\sigma}-C_{\mu\nu,\omega\tau}k_{1\rho}
k_{2\sigma}+(\rho \leftrightarrow \sigma)\biggr]
-\biggl[(\mu,\nu) \leftrightarrow (\rho,\sigma)\biggr].$

\item $I_{\mu\nu\rho\sigma\omega\tau}(k_1,k_2)=C_{\mu\nu,\rho\sigma}k_{1\tau}k_{2\omega}+\\
\Biggl\{\biggl[(C_{\sigma\omega,\nu\tau}-\eta_{\sigma\tau}\eta_{\nu\omega})
k_{1\mu}k_{2\rho}+(C_{\nu\omega,\sigma\tau}-\eta_{\sigma\omega}
\eta_{\nu\tau})k_{1\rho}k_{2\mu}+(\mu
\leftrightarrow \nu)\biggr]+ \left(\rho \leftrightarrow
\sigma\right)\Biggr\}.$
\eit

We can derive the Feynman rules from the weak gravity expansion of metric tensor on Minkowski background. The vierbein and spin connections can also be expanded accordingly. After lengthy algebraic manipulations, we can obtain the relevant Feynman rules:
\begin{center}
\begin{picture}(450,450)(0,0)

\Vertex(65,400){2.0} \ArrowLine(65,400)(35,430)
\ArrowLine(35,370)(65,400) \ArrowLine(95,431)(65,401)
\ArrowLine(95,429)(65,399) \ArrowLine(95,371)(65,401)
\ArrowLine(95,369)(65,399) \Text(42,430)[l]{$k_2$}
\Text(38,370)[l]{$k_1$}
\Text(100,431)[l]{$\mu\nu$}
\Text(100,369)[l]{$\rho\sigma$}
\Text(100,420)[l]{$p_2$}
\Text(100,360)[l]{$p_1$}
 \put(150,430){{\tiny$
i\frac{\kappa^2}{16}\{-2C_{\mu\nu,\rho\sigma}(\lks_1+\lks_2-2m)
+[(C_{\rho\sigma,\nu\alpha}-\frac{1}{4}\eta_{\nu\rho}\eta_{\sigma\alpha}-
\frac{1}{4}\eta_{\nu\sigma}\eta_{\rho\alpha})\gamma^{\alpha}(k_1+k_2)_\mu
$}}\put(150,410){{\tiny$\quad
-\frac{1}{4}(\eta_{\nu\rho}\gamma_{\sigma}+\eta_{\nu\sigma}\gamma_{\rho})(p_1-p_2)_{\mu}
+\frac{1}{4}(\lps_1-\lps_2)(\eta_{\mu\sigma}\eta_{\nu\rho}-\eta_{\nu\rho}\gamma_{\sigma}\gamma_{\mu})+(\mu\leftrightarrow\nu)]$}}
\put(150,390){{\tiny$\quad+[(C_{\mu\nu,\sigma\alpha}-\frac{1}{4}\eta_{\mu\sigma}\eta_{\nu\alpha}-
\frac{1}{4}\eta_{\nu\sigma}\eta_{\mu\alpha})\gamma^{\alpha}(k_1+k_2)_\rho
+\frac{1}{4}(\eta_{\mu\sigma}\gamma_{\nu}+\eta_{\nu\sigma}\gamma_{\mu})(p_1-p_2)_{\rho}
$}}\put(150,370){{\tiny$\quad-\frac{1}{4}(\lps_1-\lps_2)(\eta_{\mu\sigma}\eta_{\nu\rho}-
\eta_{\mu\sigma}\gamma_{\nu}\gamma_{\rho})+(\rho\leftrightarrow\sigma)]\}$}}

\Vertex(65,320){2.0} \Photon(35,350)(65,320){1.5}{5}
\ArrowLine(40,350)(60,330)\Text(53,340)[l]{$k_2$}
\ArrowLine(40,290)(60,310)\Text(53,297)[l]{$k_1$}
\Photon(35,290)(65,320){1.5}{5} \ArrowLine(115,321)(65,321)
\ArrowLine(115,319)(65,319) \Text(35,355)[b]{$\sigma$}
\Text(35,285)[l]{$\rho$} \Text(125,320)[]{$\mu\nu$}
\Text(85,310)[]{p} \put(150,320){$-i\frac{\kappa}{2}(k_1\cdot k_2
C_{\mu\nu,\rho\sigma}+D_{\mu\nu,\rho\sigma}(k_1,k_2))$}

\Vertex(65,220){2.0} \ArrowLine(65,220)(35,250)
\ArrowLine(35,190)(65,220) \Line(115,221)(65,221)
\Line(115,219)(65,219) \Text(30,240)[]{$k_2$} \Text(30,200)[]{$k_1$}
\Text(125,220)[]{$\mu\nu$} \Text(85,210)[]{p}
\put(150,220){$-i\frac{\kappa}{8}[\gamma_\mu(k_{1\nu}+k_{2\nu})
+\gamma_\nu(k_{1\mu}+k_{2\mu})
-2\eta_{\mu\nu}(\ks_1+\ks_2-2m)]$}
\Vertex(65,120){2.0} \Photon(65,120)(35,150){1.5}{6}
\ArrowLine(40,150)(55,135)\Text(48,144)[l]{$k_2$}
\Photon(35,90)(65,120){1.5}{6}
\ArrowLine(40,90)(55,105)\Text(48,96)[l]{$k_1$}

\Line(105,151)(65,121)
\Line(105,149)(65,119) \Line(105,91)(65,121) \Line(105,89)(65,119)
\Text(43,155)[l]{$\tau$} \Text(35,85)[l]{$\omega$}
\Text(115,150)[]{$\mu\nu$} \Text(120,90)[r]{$\rho\sigma$}
\Text(105,140)[]{$p_2$} \Text(105,100)[]{$p_1$}
\put(150,120){{\footnotesize$i\frac{\kappa^2}{4}(k_1\cdot k_2
C_{\mu\nu,\rho\sigma|\omega\tau}+H_{\mu\nu\rho\sigma\omega\tau}(k_1,k_2)
+I_{\mu\nu\rho\sigma\omega\tau}(k_1,k_2))$}}

\Vertex(65,40){2.0}\ArrowLine(65,40)(35,70)\Text(42,65)[l]{$k_2$}
\ArrowLine(35,10)(65,40)\Text(42,15)[l]{$k_1$}
\DashLine(65,40)(95,40){2}
\ArrowLine(65,41)(95,71)
\ArrowLine(65,39)(95,69)
\ArrowLine(65,41)(95,11)
\ArrowLine(65,39)(95,9)
\Text(100,11)[l]{$\mu\nu$}\Text(90,61)[l]{$\rho\sigma$}
\put(150,40){$i \f{y_i}{4}C_{\mu\nu,\rho\sigma}$}

\Vertex(65,-40){2.0}\ArrowLine(65,-40)(35,-10)\Text(42,-15)[l]{$k_2$}
\ArrowLine(35,-70)(65,-40)\Text(42,-65)[l]{$k_1$}
\DashLine(65,-40)(95,-10){2}
\ArrowLine(65,-41)(95,-71)
\ArrowLine(65,-39)(95,-69)
\Text(100,-61)[l]{$\mu\nu$}
\put(150,-40){$-i\f{y_i}{2}\eta^{\mu\nu}$}

\end{picture}\end{center}

\end{document}